\documentclass{aa}  

\usepackage{amssymb}
\usepackage[usenames,dvipsnames]{xcolor}
\usepackage{graphicx}
\usepackage{nicefrac}
\usepackage[rightcaption]{sidecap}
\usepackage{natbib}
\usepackage{txfonts}
\usepackage[scaled=.86]{helvet}
\usepackage[normalem]{ulem}
\usepackage{booktabs}
\usepackage{url}
\usepackage{hyperref}
\hypersetup{
    final=true,
    pageanchor=true,
    colorlinks=true,
    breaklinks=true,
    linkcolor=blue,
    citecolor=blue,
    urlcolor=blue,   
    pdfpagemode=UseNone,
    pdftitle={Sunspot Penumbra Decay},
    pdfauthor={M. Verma},
    pdfsubject={Solar Physics},
    pdfkeywords={Sun: chromosphere, Sun: photosphere, Sun: surface magnetism,
    Sun: sunspots, Techniques: image processing, Methods: data analysis}}

\usepackage{pifont}
\usepackage{epstopdf}

\newcommand\phn{\phantom{0}}

\begin{document}

%###############################################################################
%#
%#    TITLE
%#
%###############################################################################

\title{High-resolution imaging and near-infrared spectroscopy\\
    of penumbral decay}
\subtitle{} 

\author{%
    M.\ Verma\inst{1},
    C.\ Denker\inst{1},   
    H.\ Balthasar\inst{1},
    C.\ Kuckein\inst{1},
    R.\ Rezaei\inst{2},
    M.\ Sobotka\inst{3},    
    N.\ Deng\inst{4,5},
    H.\ Wang\inst{4,5},
    A.\ Tritschler\inst{6},
    M.\ Collados\inst{2},
    A.\ Diercke\inst{1,7},
    S.J.\ Gonz{\'a}lez Manrique\inst{1,8}}
    
\institute{%
    Leibniz-Institut f{\"u}r Astrophysik Potsdam (AIP),
    An der Sternwarte~16,
    14482 Potsdam,
    Germany,
%    \email{mverma@aip.de}
    \href{mailto:mverma@aip.de}{\textsf{mverma@aip.de}}
\and
    Instituto de Astrof\'{\i}sica de Canarias,
    C/ V\'{\i}a L\'actea s/n, 
    38205 La Laguna, Tenerife, Spain
\and
    Astronomical Institute, Academy of Sciences of the Czech Republic,
    Fri\v{c}ova 298,
    25165 Ond\v{r}ejov, Czech Republic
\and   
   Space Weather Research Laboratory, New Jersey Institute of Technology, 
   University Heights Newark, 
   New Jersey, USA
\and   
  Big Bear Solar Observatory, Big Bear City, 
  California, USA    
\and    
    National Solar Observatory, 
    3665 Discovery Drive
    Boulder, CO 80303, USA
\and
    Universit{\"a}t Potsdam,
    Institut f{\"u}r Physik und Astronomie,
    Karl-Liebknecht-Stra{\ss}e 24/25,
    14476 Potsdam-Golm,
    Germany
\and 
    Astronomical Institute of the Slovak Academy of Sciences, 
    05960, Tatransk\'a Lomnica, Slovakia}
\authorrunning{Verma et al.}

\date{submitted 19/08/2017, accepted 09/01/2018}

% 5 {} token are mandatory
\abstract
% % context heading (optional)
{}
% aims heading (mandatory)
{Combining high-resolution spectropolarimetric and imaging data is key to
understanding the decay process of sunspots as it allows us scrutinizing the 
velocity and magnetic fields of sunspots and their surroundings.
% three-dimensional 
}
% methods heading (mandatory)
{Active region NOAA~12597 was observed on 2016 September~24 with the 1.5-meter 
GREGOR solar telescope using high-spatial resolution imaging as well as imaging 
spectroscopy and near-infrared (NIR) spectropolarimetry. Horizontal proper 
motions were estimated with local correlation tracking, whereas line-of-sight 
(LOS) velocities were computed with spectral line fitting methods. The magnetic 
field properties were inferred with the ``Stokes Inversions based on Response 
functions'' (SIR) code for the Si\,\textsc{i} and Ca\,\textsc{i} NIR lines.
}
% results heading (mandatory)
{At the time of the GREGOR observations, the leading sunspot had two 
light-bridges indicating the onset of its decay. One of the light-bridges 
disappeared, and an elongated, dark umbral core at its edge appeared in a 
decaying penumbral sector facing the newly emerging flux. The flow and magnetic 
field properties of this penumbral sector exhibited weak Evershed flow, moat 
flow, and horizontal magnetic field. The penumbral gap adjacent to the elongated 
umbral core and the penumbra in that penumbral sector displayed LOS velocities 
similar to granulation. The separating polarities of a new flux system 
interacted with the leading and central part of the already established active 
region. As a consequence, the leading spot rotated 55$^{\circ}$ in clockwise 
direction over 12 hours. 
}
% conclusion (optional)
{In the high-resolution observations of a decaying sunspot, the penumbral 
filaments facing flux emergence site contained a darkened area resembling an 
umbral core filled with umbral dots. This umbral core had velocity and magnetic 
field properties similar to the sunspot umbra. This implies that the horizontal 
magnetic fields in the decaying penumbra became vertical as observed in 
flare-induced rapid penumbral decay, but on a very different time-scale.
}

\keywords{Sun: photosphere --
    Sun: sunspots --
    Sun: magnetic fields --
    Sun: infrared --
    Techniques: imaging spectroscopy --
    Techniques: spectroscopic}

\maketitle

%===============================================================================
%    Introduction
%===============================================================================

\section{Introduction}

% Penumbra; decay
The formation of sunspots is a rather rapid process \citep[e.g.,][]{Leka1998}
compared to their lifetime. Theory suggests that a pore develops a penumbra when
the magnetic field reaches an inclination of 45$^{\circ}$ at its edges 
\citep{Rucklidge1995, Tildesley2004}. On the other hand, penumbral decay is a 
slow process,  unless it is related to solar flares or other eruptive phenomena,
e.g., rapid penumbral decay was observed following X-class solar flares 
\citep{Wang2004, Deng2005}.  
 
The flow and magnetic fields of decaying sunspots, despite being the subject of 
many observational and theoretical studies, still lack a comprehensive physical 
description. Various aspects of sunspot decay were extensively studied 
\citep[e.g.,][]{Meyer1974, McIntosh1981, Pillet2002, Solanki2003}. Stable 
leading sunspots and irregular trailing sunspots have different decay rates 
\citep{Pillet2002}. Various decay laws are put forward to explain the decay 
rates such as a linear decay law by \citet{Bumba1963} and a parabolic decay by 
\citet{Petrovay1997}. These decay laws were reviewed by \citet{Pillet2002} 
including diffusion models to explain how magnetic flux is distributed over a 
larger area when a spot decays. However, according to the latter author, the 
flux removal process still needs a satisfactory explanation. 

% Theoretical aspect of penumbra formation and decay
\citet{Rempel2015} numerically simulated a sunspot (with penumbra) and a naked 
sunspot (penumbra removed after 20~hours) to analyze how the presence of a 
penumbra influences the moat flow and sunspot decay. He found large-scale flows 
surrounding both spots with the exception of the Evershed flow, which should 
only exist in the presence of a penumbra. In these simulations, the submergence 
of the horizontal magnetic field was the dominant decay process. A strong reduction 
of the downflow filling factor and convective rms-velocity, underneath the 
sunspot penumbra and the outer boundary of the naked spot, were the two factors 
in the simulation, which inhibited the decay process. 

% Spectropolarimetry of a Decaying Sunspot Penumbra 
\citet{BellotRubio2008} discovered finger-like structures near a sunspot in the 
last stages of evolution. These features were small-scale inhomogeneities in 
the magnetic canopy and were neither related to the penumbral filament nor to 
the Evershed flow. They speculated that these structures are related to 
penumbral field lines that no longer carry strong Evershed flows and rise to the 
chromosphere, thus producing the disappearance of the penumbra at photospheric 
levels. Moving magnetic features (MMFs) are present around many 
decaying sunspots \citep[e.g.,][]{Harvey1973}. \citet{Verma2012a} noted that 
flux carried by MMFs from the decaying spot reaches the surrounding 
supergranular boundary, similar to what is described by \citet{Deng2007}. 
However, the relation between penumbral filaments, MMFs, Evershed flow, and 
moat flow is still a matter of debate \citep[e.g.,][]{Dalda2005, 
CabreraSolana2006}. 

% Magnetic and flow fields in formation and decay of sunspot
% Observations of formation of penumbra in emerging flux region
The changes in the size and the magnetic/velocity field of umbra and penumbra 
can give some indication, when sunspots begin to form or decay. Around a 
decaying sunspot, \citet{Balthasar2013} reported a reversed gradient in the 
vertical magnetic field component in the penumbra. The authors interpret it as 
an inclination affect, which is already seen in \citet{Balthasar2008}. This 
height dependence was derived by comparing the magnetic field obtained from the 
infrared lines Fe\,\textsc{i} 1078.3~nm and Si\,\textsc{i} 1078.6~nm. 
\citet{Watanabe2014} observed formation and decay of a rudimentary penumbra in a 
protospot \citep{Leka1998}, where the penumbra developed at the expense of 
umbral magnetic flux. While the penumbra decayed, they observed that the 
penumbral field became vertical resulting in the recovery of umbral area, i.e., 
penumbral decay leads to rearrangement of magnetic field lines. In contrast, 
\citet{Schlichenmaier2010b} noticed that the umbral area remained constant when 
the sunspot was developing a penumbra. Based on a study of ten sunspots, 
\citet{Jurcak2011} concluded that the inner penumbral boundaries are defined by 
the critical value of the vertical component of the magnetic field, i.e., 
$B_\mathrm{stable} = 1.8$~kG. Recently, \citet{Jurcak2017} studied penumbra 
formation around a pore, which supports the scenario earlier proposed by 
\citet{Jurcak2011} and \citet{Jurcak2015}, i.e., the stable vertical component 
of magnetic field is needed to establish the umbra-penumbra boundary.

% What we show here
The magnetic field in the vicinity of sunspots also plays a significant role in 
formation and decay of sunspot penumbra. \citet{Kuenzel1969} found that the 
polarity, magnetic field strength, and size of a neighboring spot influence the 
penumbra of the sunspot. He also noticed that no penumbra forms between the 
spots of same polarity. Flux emergence in the vicinity of a sunspot inhibits the 
formation of a stable penumbra as demonstrated by \citet{Schlichenmaier2010b} 
and \citet{Rezaei2012}. Both studies agree that stable penumbral filaments only 
form away from the flux emergence site, implying a ``quiet'' environment as a 
prerequisite for developing a penumbra. \citet{Lim2013} observed the formation 
of a non-radial penumbra in a flux-emergence region with pre-existing 
chromospheric canopy fields. They suggested that in an emerging region, the 
penumbra is formed when the emerging flux is constrained from developing any 
further by the overlying chromospheric canopy fields. However, 
\citet{Murabito2017} observed that a stable penumbra was formed in a sunspot at 
the site facing the opposite polarity and the region of flux emergence. These 
sometimes contradictory findings lead to the question what role does magnetic 
flux emergence and its interaction with the overlying canopy play in the decay 
of sunspot penumbrae? The present work contributes to answering this still open 
question. We present high-resolution GREGOR observations of a decaying sunspot, 
where the decaying penumbra faces the flux emergence site.

%-------------------------------------------------------------------------------
%    Figure 1: Blue continuum image from GREGOR
%-------------------------------------------------------------------------------
\begin{figure}[t]
\includegraphics[width=\columnwidth]{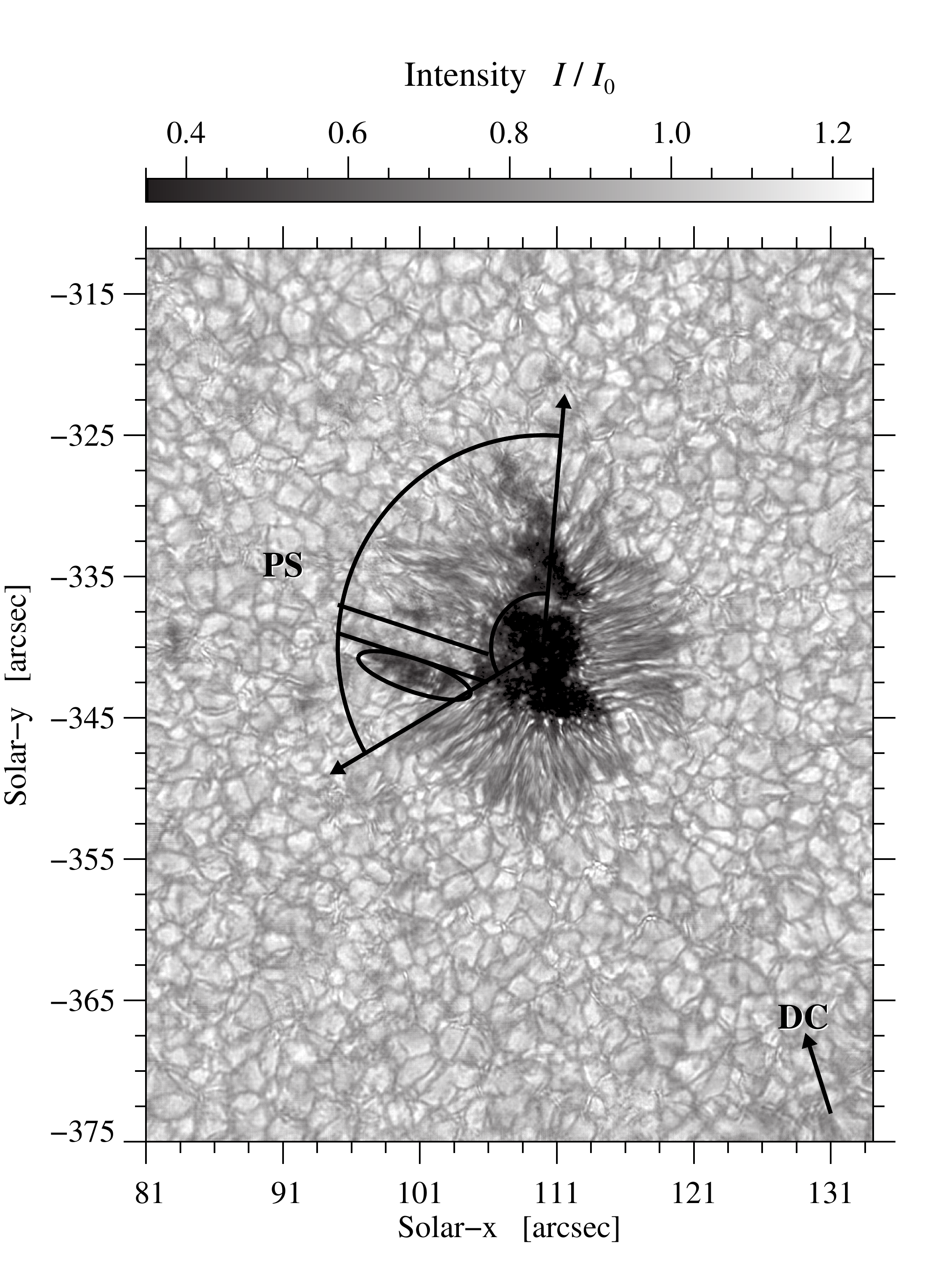}
\caption{Speckle-restored blue continuum image from HiFI observed at 08:50~UT on 
    2016 September~24. The two black arcs mark the penumbral sector \textsf{PS} 
    discussed in Sect.~\ref{SEC03}. The arrow in the lower right corner points 
    to solar disk center \textsf{DC}. The black oval and the two parallel lines 
    inside the \textsf{PS} indicate the elongated, dark umbral core and the 
    penumbral gap, respectively, which are discussed in Sect.~\ref{SEC03}.}
\label{FIG01}
\end{figure}

%===============================================================================
%    Observations
%===============================================================================

\section{Observations and data reduction\label{SEC02}}

%-------------------------------------------------------------------------------
%    Observations
%-------------------------------------------------------------------------------

\subsection{Observations}

The aim of 2016 September coordinated observing campaign was to obtain 
high-resolution spectropolarimetric data and accurate measurements of the 
photospheric and chromospheric three-dimensional magnetic and flow fields within 
an active region. The campaign utilized the following telescopes and 
instruments: the GREGOR Fabry-P{\'e}rot Interferometer \citep[GFPI,][and 
references therein]{Puschmann2012, Denker2010b}, the High-resolution Fast Imager 
\citep[HiFI,][]{Kuckein2017}, and the GREGOR Infrared Spectrograph 
\citep[GRIS,][]{Collados2012} at the 1.5-meter GREGOR solar telescope 
\citep{Schmidt2012, Denker2012}; the Echelle Spectrograph at the Vacuum Tower 
Telescope \citep[VTT,][]{vonderLuehe1998}; the Spectro-Polarimeter 
\citep[SP,][]{Ichimoto2008} of the Solar Optical Telescope 
\citep[SOT,][]{Tsuneta2008} on board the Japanese space mission Hinode 
\citep{Kosugi2007}; and the Helioseismic and Magnetic Imager 
\citep[HMI,][]{Scherrer2012, Schou2012} and  the Atmospheric Imaging Assembly 
\citep [AIA,][]{Lemen2012} on board the Solar Dynamics Observatory 
\citep[SDO,][]{Pesnell2012}. The observations were carried out for 10 days. This 
study is based on just one of the datasets containing a decaying sunspot, which 
was observed with GFPI, HiFI, and GRIS. Continuum images and magnetograms of HMI 
provide context information.

% Information about active region
Active region NOAA~12597 emerged on the solar surface on 2016 September~22. It 
was classified as a simple bipolar $\beta$-region with a strong leading spot and 
scattered pores forming the trailing part. By 00:00~UT on 2016 September~24, the 
leading sunspot was encircled by a complete penumbra. The trailing part 
contained two large pores. The GREGOR observations started two days later on 
2016 September~24. The region already crossed the central meridian and was 
located at coordinates (110\arcsec, $-350$\arcsec) corresponding to $\mu \approx 
0.92$ on the solar disk. During the observations the mature leading sunspot 
started to decay. The region was relatively quiet and only two B-class solar 
flares occurred, one before and one after the GREGOR observations. 

% Observations from GFPI
The leading sunspot of the region was observed with the GFPI in spectroscopic 
mode starting at 08:52~UT on 2016 September~24. The instrument scanned the 
photospheric Fe\,\textsc{i} $\lambda 617.3$~nm line. The imaging spectrometer 
recorded in total 140 scans with 25 equidistantly spaced wavelength points. One 
spectral scan took about 25~s with an exposure time of 10~ms and eight frames at 
each wavelength point. In addition, strictly simultaneous broad-band images were 
captured using an interference filter with a full-width-at-half-maximum (FWHM) 
of 10~nm  centered at $\lambda$612.3~nm. Broad- and narrow-band images have $688 
\times 512$ pixels after 2$\times$2-pixel binning with an effective image scale 
of about 0.081\arcsec\ pixel$^{-1}$ resulting in a field-of-view (FOV) of about 
$56\arcsec \times 41\arcsec$.

% Observations from HiFI 
The two HiFI cameras recorded strictly simultaneous high-resolution G-band 
$\lambda$430.7~nm and blue continuum $\lambda$450.5~nm images with an exposure 
time of 1.8~ms. Initially, sets of 500 frames with a frame rate of 47~Hz were 
acquired but ultimately, after frame selection, only the best 100 calibrated 
frames were stored for post-processing and image restoration. The sCMOS imagers 
have a pixel size of 6.5~$\mu$m $\times$ 6.5~$\mu$m. The images contain 2560 
$\times$ 2160 pixels, and the image scale is about 0.025\arcsec\ pixel$^{-1}$, 
which leads to a FOV of about $65\arcsec \times 55\arcsec$. The total duration 
of the HiFI time-series was about 46~min. Only HiFI blue continuum images are 
used in this study to provide high-resolution context information.

% Observations from GRIS
In parallel to GFPI and HiFI, we observed in the He\,\textsc{i} 
$\lambda$1083.0~nm spectral region with GRIS recording spectropolarimetric data. 
This NIR spectral window contains various solar and telluric lines. However, of 
primary interest are the photospheric Si\,\textsc{i} $\lambda$1082.7~nm line, 
the chromospheric He\,\textsc{i} $\lambda$1083.0~nm triplet (two blended lines 
$\lambda$1083.030~nm and $\lambda$1083.025~nm), and the photospheric 
Ca\,\textsc{i} $\lambda$1083.9~nm line. The slit-width was about 0.25\arcsec. 
The number of spectral points was $n=1010$, and the spectral sampling was 
1.81~pm pixel$^{-1}$. The scan covered 51.8\arcsec\ in 360 steps with a step 
size of about 0.144\arcsec. The image scale along the slit was about 
0.136\arcsec\ pixel$^{-1}$ resulting in a FOV of about $62\arcsec \times 
52\arcsec$. The integration time used for each slit position was 100~ms with 10 
accumulations. Thus, the first scan of the sunspot with 360 steps took 36~min, 
and the second with 300 steps finished in 30~min. Only the first scan starting 
at 09:02~UT is analyzed in detail, while the second scan starting at 10:30~UT 
provides supporting context information. Since, the installation of three-mirror 
system as de-rorator in the GREGOR, the observed images and spectra do not have 
to be de-rotated any longer and can easily be matched with orientation of the 
SDO images.

%-------------------------------------------------------------------------------
%    Data reduction and analysis
%-------------------------------------------------------------------------------

\subsection{Data reduction and analysis \label{SEC02.2}}

% \subsubsection{GFPI and HiFI \label{SEC02.2.1}}
Calibration of GFPI and HiFI data was carried out with the data processing 
pipeline ``sTools'' \citep{Kuckein2017}. The sTools pipeline performs 
basic data calibration such as dark-frame subtraction and flat-field 
corrections for both post-focus instruments. In addition, for GFPI data it 
computes and applies prefilter transmission and blueshift corrections along 
with removing the tilt in the spectral profile by matching it with the Fourier
Transform Spectrometer \citep[FTS,][]{Neckel1984} spectral atlas. After basic 
calibration, sTools also performs the image restorations providing us with 
Level~2 data. Narrow- and broad-band images were restored using 
Multi-Object Multi-Frame Blind Deconvolution \citep[MOMFBD,][]{Loefdahl2002, 
vanNoort2005}. The Kiepenheuer Institute speckle interferometry package 
\citep[KISIP,][]{Woeger2008a, Woeger2008b} was used to restore the blue 
continuum and G-band images. For both GFPI and HiFI images, the image contrast, 
degraded by scattered light, was then corrected using the method described in 
\citet{bello2008}. One example of a restored blue continuum image is 
shown in Fig.~\ref{FIG01}. The restored HiFI images are finally written as
image extensions in the Flexible Image Transport System \citep{Wells1981,
Hanisch2001} format  for further use and analysis. The analysis and 
management for GREGOR data, specifically for image restoration and imaging
spectroscopy, is described in detail in \citet{Denker2017b}.

%-------------------------------------------------------------------------------
%    Figure 2: Overview figure from SDO
%-------------------------------------------------------------------------------
\begin{figure}[t]
\includegraphics[width=\columnwidth]{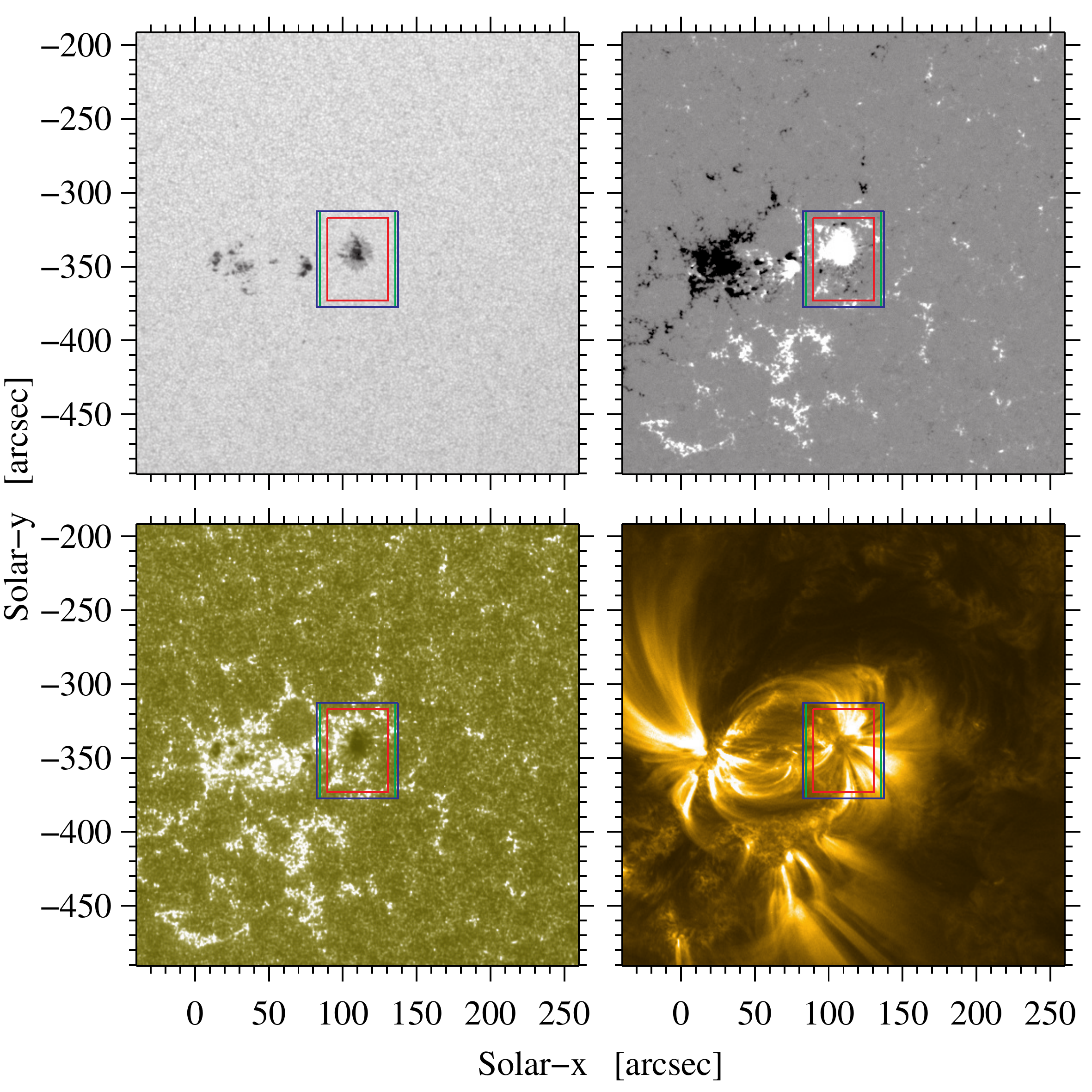}
\caption{Overview of active region NOAA 12597: HMI continuum image
    (\textit{top-left}), HMI magnetogram (\textit{top-right}), AIA
    $\lambda$160~nm image (\textit{bottom-left}), and AIA Fe\,\textsc{ix}
    $\lambda$17.1~nm image (\textit{bottom-right}) observed at 09:00~UT on 
    2016 September~24. The red, green, and blue boxes represent the FOV of 
    GFPI, GRIS scan, and HiFI images, respectively. The magnetogram was
    clipped between $\pm$250 G and the AIA $\lambda$17.1~nm image is displayed
    on a linear scale.}
\label{FIG02}
\end{figure} 

The GFPI spectra of the Fe\,\textsc{i} $\lambda$617.3~nm line provided us 
with many spectral line characteristics including FWHM, equivalent width, and 
line-shift or LOS velocity. The most important quantity is the LOS velocity
in and around the sunspot. We determined the velocity with a Fourier phase 
method \citep{Schmidt1999}. We chose this method as it utilizes the entire 
line profile, takes into consideration the spectral line asymmetry, and is 
less sensitive to noise. The line shifts calculated with this method were 
then converted into velocities using the Doppler formula. We used the average
photospheric velocity of quiet-Sun granulation as the frame of reference. In 
the computed LOS velocity map redshifts 
are positive and blueshifts are negative.

% Observations from SDO
The synoptic nature and good time coverage of SDO data complements the 
high-resolution observations and provides context information. The HMI 
continuum image and the LOS magnetogram, along with AIA EUV images in
Fe\,\textsc{ix} $\lambda$17.1~nm and FUV images at $\lambda$160~nm 
present an overview of the observed active region (Fig.~\ref{FIG02}). The 
FOVs covered by various instruments are indicated by colored rectangular 
boxes. In addition, we used continuum images and LOS magnetograms for 
every hour on September~24 to study the temporal evolution of the region.

One of the advantages of having time-series of high-resolution images at the
same time as spectroscopic data is that we can determine horizontal proper
motions as well as LOS velocities. The horizontal proper motions were computed
from the restored GFPI broad-band images, where we applied our implementation
\citep{Verma2011, Verma2013} of the local correlation tracking
\citep[LCT,][]{November1988} algorithm. The images in both time-series were
aligned with sub-pixel accuracy with respect to average image of the 
time-series and the signature of five-minute oscillations was removed using a
subsonic Fourier filter with the photospheric sound speed as cut-off velocity.
The LCT velocities were computed over the image tiles of 48 $\times$ 48 pixels
with a Gaussian kernel having a FWHM = 1200~km. Consecutive images were used 
to calculate LCT maps, i.e., the cadence was $\Delta t = 25$~s, and the 
computed maps were then averaged over a time period of $\Delta T=40$~min.

% \subsubsection{GRIS \label{SEC02.2.2}}

The basic data reduction for GRIS is carried out on-site using the GRIS data 
pipeline (Collados, priv.\ communication). The various steps include dark-frame 
subtraction, flat-field correction, crosstalk removal, and calibration of the 
polarization modulator \citep{Collados1999}. Details of the next data reduction 
steps such as wavelength calibration and correction for the spectrograph 
profile, etc. are given in \citet{Verma2016b}. We used the ``Stokes Inversion 
based on Response functions'' \citep[SIR,][]{RuizCobo1992} code to invert the 
GRIS spectra of the two scans. We restricted ourselves to the photospheric lines 
Si\,\textsc{i} $\lambda$1082.7~nm ($g_\mathrm{eff} = 1.5$) and Ca\,\textsc{i} 
$\lambda$1083.9~nm ($g_\mathrm{eff} = 1.5$). The inversion for two lines were 
performed separately because velocity and magnetic field were implemented 
independent of height. The starting model for Si\,\textsc{i} covered the optical 
depth range $+1.0 \le \log \tau \le -5.4$, whereas for Ca\,\textsc{i} the 
optical depth range was $+1.0 \le \log \tau \le -4.4$. For both scans a 
limb-darkening factor was considered according to Eq.~(10) of 
\citet{Pierce1977b}. We assumed a constant macroturbulence of 1~km~s$^{-1}$ and 
2~km~s$^{-1}$ for the Si\,\textsc{i} and Ca\,\textsc{i} lines, respectively. In 
addition, for both lines a fixed stray light contribution of two percent was 
used. This value was chosen as a lower limit \citep{Balthasar2016}. The 
inversions for both lines delivered the temperature stratification with three 
nodes $T(\tau)$, the total magnetic flux density $B_\mathrm{tot}$, the magnetic 
inclination $\gamma$ and azimuth $\phi$, and the Doppler velocity 
$v_\mathrm{LOS}$ that is constant with height (one node for each of these 
physical parameters). The magnetic azimuth ambiguity for the inversions was 
solved by assuming the center of leading spot as the azimuth center 
\citep{Balthasar2006}. Finally, we rotated the magnetic vector with respect to 
the local solar frame. Pixels for which the circular and the linear polarization 
were below 0.2\% degree of polarization, were excluded from any further analysis 
of the magnetic field, as this value corresponds to the noise level of the 
polarization signal.

In addition to the SIR results, the LOS velocities and line-core intensities for 
the above mentioned Si\,\textsc{i}, Ca\,\textsc{i}, and He\,\textsc{i} lines 
were derived using a single Lorentzian model to fit the line core 
\citep{GonzalezManrique2016}. The model is given by $L = a_0 / (u^2 + 
1) + a_3$ with $u = (x-a_1) / a_2$, where $a_0$ is the amplitude, $a_1$ is the 
peak centroid, $2a_2$ is the FWHM, and $a_3$ is a constant representing the 
continuum. All the shown GRIS maps are trimmed and aligned to match the GFPI's 
FOV.

When computing the LOS velocities for all spectral lines, we considered 
quiet-Sun granulation as the frame of reference. The reference for the 
chromospheric He\,\textsc{i} line, was a quiet-Sun region, where additionally 
filamentary structures were absent. Because the photospheric lines exhibit 
convective blueshifts, we computed the velocity offsets based on the SIR results 
with respect to the umbra, which can be assumed at rest. The approximate 
velocity offsets are Fe\,\textsc{i} $-215$~m~s$^{-1}$, Ca\,\textsc{i} 
$-300$~m~s$^{-1}$, and Si\,\textsc{i} $-50$~m~s$^{-1}$, which slightly depend on 
the selected part of the umbra. Here, we chose the entire umbra encircled by the 
black contours. However, other choices, for example the darkest part of the 
umbra, affected the offsets by only about 50~m~s$^{-1}$. As a result, the umbra 
appears redshifted in all photospheric LOS velocity maps but the fine structure 
of penumbra and granulation is easier to compare across different spectral line 
in this way. Our results agree with the trend for the convective blueshift as a 
function of line depth as given by, for example, \citet{Reiners2016}.

%===============================================================================
%    Results
%===============================================================================

\section{Results\label{SEC03}}

%-------------------------------------------------------------------------------
%    Temporal evolution of the active region
%-------------------------------------------------------------------------------

\subsection{Temporal evolution of the active region}

%-------------------------------------------------------------------------------
%    Figure 3a: Overview figure from SDO
%-------------------------------------------------------------------------------
\begin{figure*}[t]
\includegraphics[width=\textwidth]{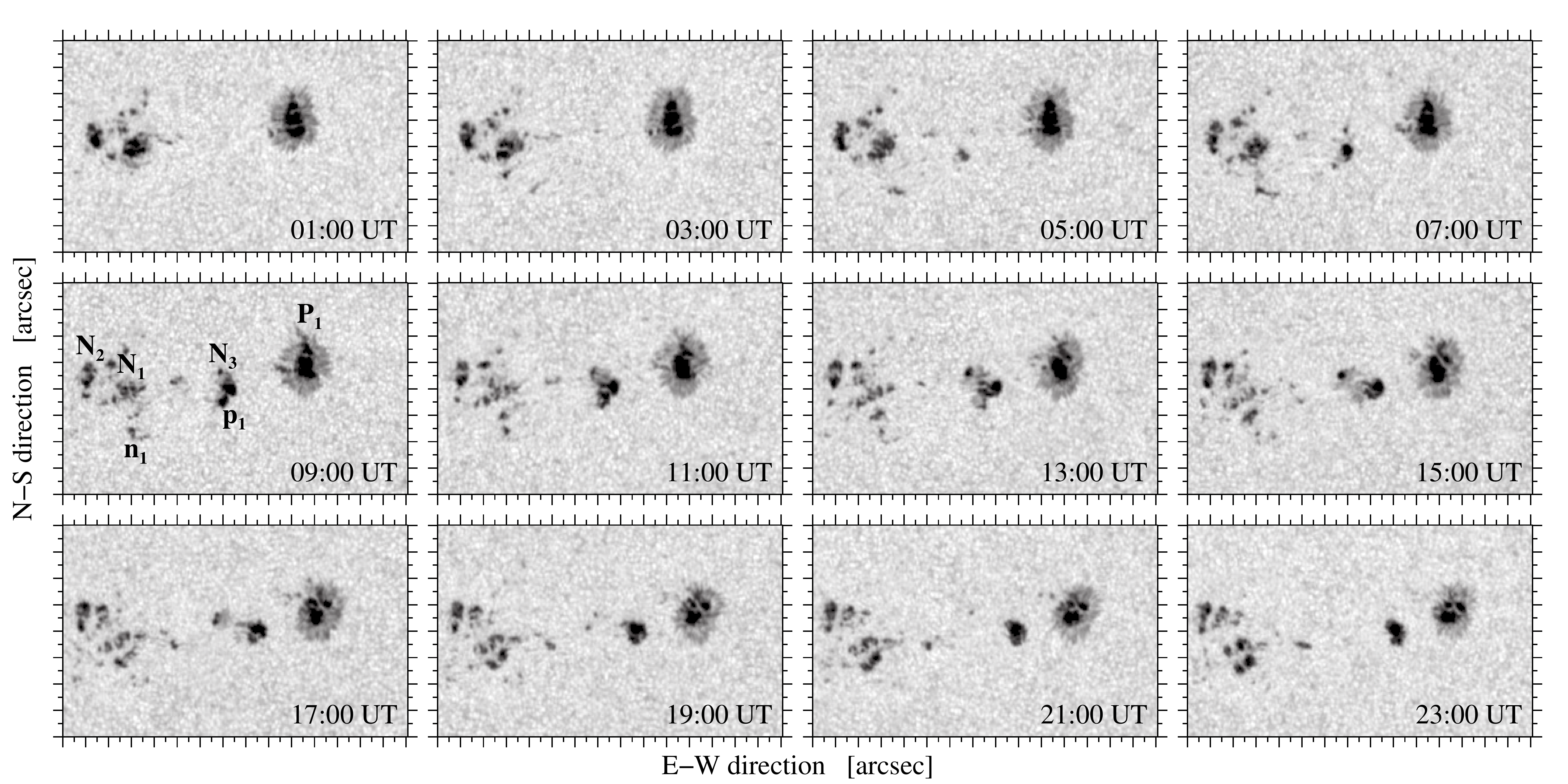}
\caption{Evolution of active region NOAA~12597 on 2016 September~24 based on
    HMI continuum images with a FOV of $150\arcsec \times 80\arcsec$. Major
    tick marks are placed at 10\arcsec-intervals. The
    images were displayed at two-hour intervals starting at 01:00~UT. The
    abbreviations in the 09:00~UT panel indicate the leading spot
    \textsf{P$_\mathsf{1}$}, the trailing pores \textsf{N$_\mathsf{1}$} and
    \textsf{N$_\mathsf{2}$}, the central negative pore \textsf{N$_\mathsf{3}$},
    and the leading and trailing pores \textsf{p$_\mathsf{1}$} and
    \textsf{n$_\mathsf{1}$} of the emerging active region.}
\label{FIG03a}
\end{figure*} 

%-------------------------------------------------------------------------------
%    Figure 3b: Overview figure from SDO
%-------------------------------------------------------------------------------
\begin{figure*}[t]
\includegraphics[width=\textwidth]{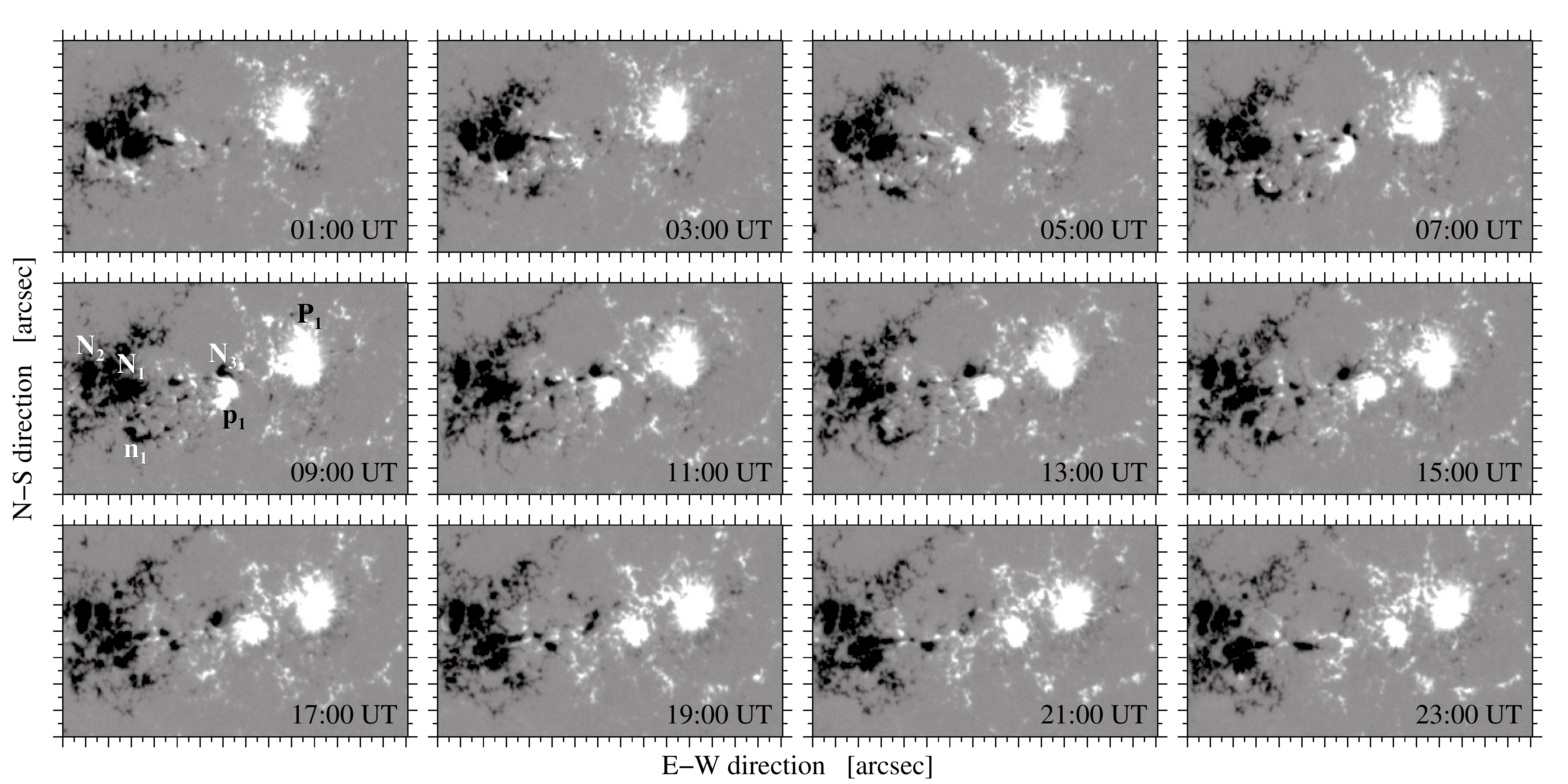}
\caption{Evolution of active region NOAA~12597 on 2016 September~24 based on 
    HMI magnetograms, which were scaled between $\pm 500$~G. Annotations, 
    labels, and FOV are the same as in Fig.~\ref{FIG03a}.}
\label{FIG03b}
\end{figure*}

% The phase of new flux emergence
Studying the evolution of active region NOAA~12597 on 2016 September~24 is based 
on HMI continuum images and magnetograms along with HiFI images. On the day of 
the GREGOR observations, the sunspot group was a simple bipolar region with a 
weak negative-polarity trailing part and a positive-polarity leading sunspot. 
The observations focused on the leading mature sunspot. Figures~\ref{FIG03a} 
and~\ref{FIG03b} contain 150\arcsec$\times$80\arcsec-snapshots of the region, 
where the continuum images and LOS magnetograms are displayed at two-hour 
intervals starting from 01:00~UT on 2016 September~24. The whole region can be 
divided into two flux systems: a large decaying active region (DAR) and a small 
emerging active region (EAR) to the south of the DAR's trailing part. The 
relevant spots and pores are labeled in the 09:00~UT panel of Figs.~\ref{FIG03a} 
and~\ref{FIG03b} to clearly identify them for the subsequent analysis and 
discussion, i.e., the main leading spot \textsf{P$_\mathsf{1}$}, the trailing 
pores \textsf{N$_\mathsf{1}$} and \textsf{N$_\mathsf{2}$}, the central negative 
pore \textsf{N$_\mathsf{3}$}, the leading pore \textsf{p$_\mathsf{1}$}, and the 
trailing pore \textsf{n$_\mathsf{1}$} in the EAR. The nomenclature incorporates 
both magnetic polarity and membership with respect to DAR and EAR. 
The membership to DAR and EAR was assigned based on the 
evloution of the region seen in the time-lapse movies of SDO. Since the 
indication of \textsf{N$_\mathsf{3}$} was already there in the magnetogram at 
01:00~UT before the EAR started to develop, we ascribed it to the DAR.

At 01:00~UT, \textsf{P$_\mathsf{1}$} displayed two light-bridges indicating the 
onset of its fragmentation. In the next two hours at 03:00~UT, 
\textsf{P$_\mathsf{1}$} remained unchanged. At 05:00~UT, the penumbral sector 
\textsf{PS} (marked in Fig.~\ref{FIG02}) of \textsf{P$_\mathsf{1}$}, which faced 
the trailing region, started to decay. A gap appeared at the edge of lower 
light-bridge of \textsf{P$_\mathsf{1}$}. At 07:00~UT, this gap resembled umbral 
cores and stretched out towards the trailing polarity in the magnetogram. In 
addition, the lower light-bridge in \textsf{P$_\mathsf{1}$} disappeared. At 
09:00~UT, the GREGOR observations started. The spot \textsf{P$_\mathsf{1}$} now 
contained two elongated features like umbral cores within \textsf{PS} -- one on 
top and the other on the side facing the trailing part. Only the upper 
light-bridge remained. The high-resolution HiFI images revealed further details 
within \textsf{P$_\mathsf{1}$} (Fig.~\ref{FIG02}). Both G-band and 
blue-continuum images exhibited umbral dots at the location of the disappearing 
light-bridge. The dark extrusion, which resembled an umbral core, was located at 
the edge of this light-bridge (black oval in Fig.~\ref{FIG02}). In the gap, 
between the umbral core and the penumbral sector (indicated in 
Fig.~\ref{FIG02} by two parallel lines), granulation was present. The upper 
light-bridge started to fork creating a \textsf{Y}-shape. The G-band images 
reveal many G-band bright points in the surroundings indicative of small-scale 
flux elements. The time-series of GFPI broad-band images uncover many minutes 
and fast changes in the spot, which are discussed in Sect.~\ref{SEC03.2}. At 
11:00~UT, \textsf{P$_\mathsf{1}$} exhibited clear signs of penumbral decay. 
Comparing earlier images with those at 13:00~UT, \textsf{P$_\mathsf{1}$} 
distinctly rotated, which is evident when examining position and direction of 
the light-bridge. In the four hours from 09:00--13:00~UT, the sunspot rotated 
30$^{\circ}$ clockwise with a rotation rate of about 15$^{\circ}$ every two 
hours. At this time, \textsf{PS} almost decayed. This process continued for the 
next two hours as seen in the continuum images and magnetograms at 15:00~UT. The 
rotation of \textsf{P$_\mathsf{1}$} continued and became more prominent. By 
17:00~UT, the rotation of \textsf{P$_\mathsf{1}$} was very obvious, and new 
light-bridges appeared. However, the rotation rate had slowed down. In the last 
four hours of the day, \textsf{P$_\mathsf{1}$} became more compact, i.e., with a 
decreasing penumbral coverage, with stronger light-bridges, and with substantial 
rotation of almost 55$^{\circ}$ in clockwise direction with respect to the image 
at 09:00~UT.

In the beginning at 01:00~UT, the trailing part consisted of two big pores with 
\textsf{N$_\mathsf{1}$} on the south having a rudimentary penumbra and 
\textsf{N$_\mathsf{2}$} on the north, along with some scattered pores. The 
region between the two main polarities contained only granulation. However, in 
the corresponding magnetogram, small-scale mixed polarities were present at that 
location. In next two hours at 03:00~UT, \textsf{N$_\mathsf{1}$} developed a 
light-bridge. Over the next hours, continuum images revealed that the trailing 
pores started to disintegrate. By the end of the day, the trailing part of the 
active region shrunk with few small scattered pores remaining.

The central part of the active region contained only granulation at 01:00~UT. 
Some small-scale mixed-polarity patches near the trailing pores were still too 
weak to leave an imprint in the continuum images. The central negative pore 
\textsf{N$_\mathsf{3}$} belonging to DAR was not visible in the continuum images 
until 05:00~UT. However, its presence was already noticeable in the magnetograms 
at 03:00~UT. Throughout the day the central pore mainly interacted with the 
positive polarity of the EAR, which appeared as separate magnetic flux-bundle 
rising onto the surface at the lower right-side of \textsf{N$_\mathsf{1}$}. 
Signs of flux emergence were present in the 03:00~UT magnetograms, where the 
infant stages of two polarities were first evident. The leading and trailing 
part of the EAR developed at 05:00~UT as positive \textsf{p$_{1}$} and negative 
\textsf{n$_\mathsf{1}$} polarity pores, respectively. At 07:00~UT, the center of 
the EAR contained many mixed polarity features. In addition, as often seen in 
flux-emergence regions, \textsf{p$_\mathsf{1}$} separated and moved away from 
\textsf{n$_\mathsf{1}$}. In this process, \textsf{p$_\mathsf{1}$} moved closer 
to \textsf{N$_\mathsf{3}$}. By 09:00~UT, \textsf{p$_\mathsf{1}$} grew in size by 
coalescence of small-scale magnetic features while \textsf{N$_\mathsf{3}$} 
remained in close proximity. At 11:00~UT, both magnetograms and continuum images 
showed that \textsf{p$_\mathsf{1}$} and \textsf{N$_\mathsf{3}$} slid past each 
other like two cars passing each other along the central divider on a highway. 
This motion had an immediate affect on \textsf{P$_\mathsf{1}$} and resulted in 
the clockwise rotation of the whole leading sunspot. 

In addition, the opposite-polarity pores \textsf{p$_\mathsf{1}$} and 
\textsf{n$_\mathsf{1}$} were still separating and moving away from each other. 
At 13:00~UT, the most dynamic location was still the center of the FOV. There, 
\textsf{N$_\mathsf{3}$} moved towards the trailing part and 
\textsf{p$_\mathsf{1}$} towards the leading part of the DAR, which continued at 
13:00~UT. The declining phase of the EAR began at 17:00~UT. The central pores 
\textsf{p$_\mathsf{1}$} and \textsf{N$_\mathsf{3}$} were clearly separated. The 
trailing part of the EAR already decayed. Within the next two hours, 
\textsf{p$_\mathsf{1}$} and \textsf{N$_\mathsf{3}$} moved closer to the leading 
and trailing parts, respectively. Although, now only faint signatures of 
\textsf{N$_\mathsf{3}$} were visible in continuum images. Mixed polarities were 
still present in the region of flux emergence, but the overall photometric 
morphology and magnetic topology significantly simplified.

We used AIA $\lambda$171~nm EUV images to examine the topology of the coronal 
loop structure related to the active region. Early on 2016 September~24, the 
loops had a bipolar active region structure, i.e., with arches connecting the 
two opposite polarities of the DAR, which were a little lopsided towards the 
leading sunspot \textsf{P$_\mathsf{1}$}. Onwards 04:00~UT, with the emergence of 
new flux, a low-lying loop structure connecting \textsf{p$_\mathsf{1}$} and 
\textsf{n$_\mathsf{1}$} was established at its location, disrupting the DAR loop 
structure. The flux emergence caused some small-scale brightenings and a small 
disruption in the large overarching loops. However, after the initial flux 
emergence phase the usual large bipolar arches reestablished themselves. The 
rotation of \textsf{P$_\mathsf{1}$} in photospheric layers caused the rotation 
of this foot point. However, the trailing polarity footpoint remained stable. In 
addition, loops terminating at \textsf{P$_\mathsf{1}$} expanded, slightly 
rotated, fanned out,  and detached towards the western limb. 

%-------------------------------------------------------------------------------
%    Figure 4: pictorial depiction
%-------------------------------------------------------------------------------
\begin{figure}
\center
\includegraphics[width=\columnwidth]{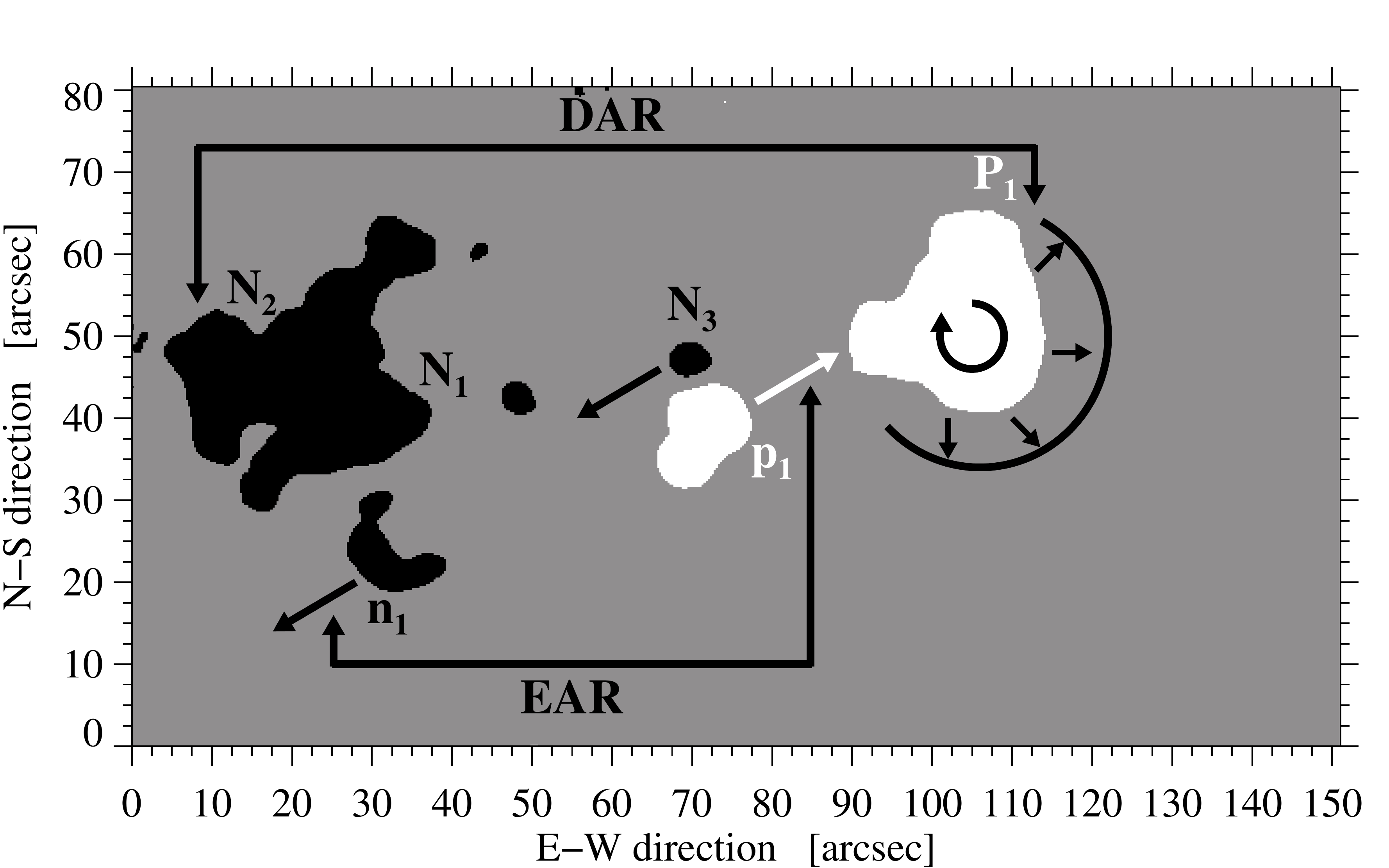}
\caption{The sketch is based on the HMI magnetogram observed at 09:00~UT on 
    2016 September~24. Black and white areas mark the negative and positive 
    polarities, respectively. Arrows indicate the direction in which different 
    features moved over the course of the day. The black semi-circle denotes 
    the region around \textsf{P}$_\mathrm{1}$ with moat flow and MMFs. The 
    black lines with  vertical arrows mark the extent and features belonging to 
    DAR and EAR, also labeled with upper- and lowercase letters.}
\label{FIG04}
\end{figure}

We summarized the main dynamical properties of the region in the sketch depicted 
in Fig.~\ref{FIG04}. The sketch is based on the magnetogram at 09:00~UT, which 
was smoothed and clipped at $\pm 250$~G. All features are labeled as in 
Figs.~\ref{FIG03a} and~\ref{FIG03b}. The main dynamics within the active region 
were indicated by arrows, i.e., the separation of \textsf{p$_\mathsf{1}$} and 
\textsf{n$_\mathsf{1}$}, the sliding of \textsf{N$_\mathsf{3}$} and 
\textsf{p$_\mathsf{1}$}, and the resulting clockwise rotation of 
\textsf{P$_\mathsf{1}$}. The arrows only show the direction but not the 
magnitude of these motions.

\begin{figure}
\includegraphics[width=\columnwidth]{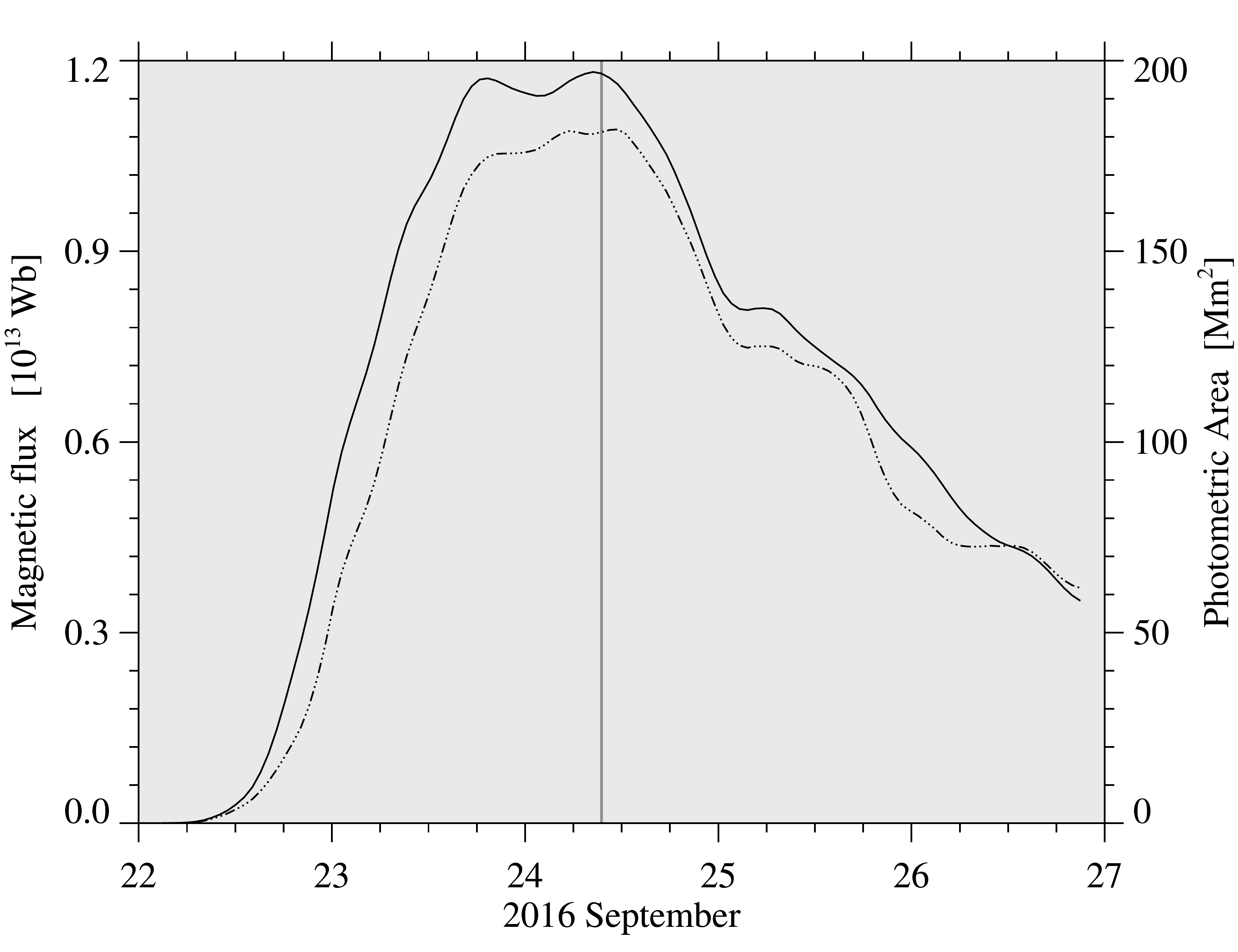}
\caption{Temporal evolution of the area (dash-dotted) and 
    magnetic flux (solid) of the leading spot \textsf{P$_\mathsf{1}$} using
    HMI continuum images and LOS magnetograms. The vertical 
    line indicates the start of GREGOR observations.}
\label{FIG011}
\end{figure}

To quantify the overall decay rate of the leading spot, we followed the active 
region for five days and found that the region had recurring flux emergence at 
the same location. There were  continuously merging flux elements in the 
trailing part. However, only the leading spot of the DAR, i.e., 
\textsf{P$_\mathsf{1}$} remained separated and did not merge with the 
neighboring emerging flux system. Hence, we computed flux and area growth and 
decay rates for only \textsf{P$_\mathsf{1}$} covering five days, i.e., September 
22--27. We created a binary template based on HMI LOS magnetograms that 
contained only \textsf{P$_\mathsf{1}$}. We used a flux threshold of 500~G along 
with morphological erosion using a 1-Mm kernel, followed by dilation using a 
5-Mm kernel. To determine the photometric area, we used contrast-enhanced HMI 
continuum images. The sunspot was identified with the help of intensity 
thresholding, i.e., \textcolor{Black}{$I_\mathrm{spot} < 0.95 I_{0}$}, where 
$I_{0}$ refers to the normalized quiet-Sun intensity. The measured area and 
magnetic field strength are taken at face value. We used linear regression to 
compute the growth and decay rates, because neither magnetic flux nor 
photometric area exhibited any indication for a parabolic growth or decay law. 
As seen in Fig.~\ref{FIG011}, the sunspot grew faster than it decayed. GREGOR 
observations (grey line) marked the start of the decay process. The flux decay 
rate of \textcolor{Black}{$0.33 \times 10^{13}$~Wb~day$^{-1}$ was almost} two 
and a half times lower than the growth rate of \textcolor{Black}{$0.80 \times 
10^{13}$~Wb~day$^{-1}$}. The growth and decay rates for the leading spot 
\textsf{P$_\mathsf{1}$} agree with earlier results 
\citep[e.g.,][]{MartinezPillet2002, Otsuji2011, Kubo2008a}. Similarly, the 
photometric area decay rate of \textcolor{Black}{50~Mm$^2$~day$^{-1}$} was also 
\textcolor{Black}{about} two and a half times lower than the growth rate of 
\textcolor{Black}{110~Mm$^2$~day$^{-1}$}.

%-------------------------------------------------------------------------------
%     Figure 11: Magnetic flux and Area evolution from SDO 
%-------------------------------------------------------------------------------

%-------------------------------------------------------------------------------
%    Horizontal velocities
%-------------------------------------------------------------------------------

\subsection{Horizontal proper motions\label{SEC03.2}}

\begin{figure*}
\includegraphics[width=\textwidth]{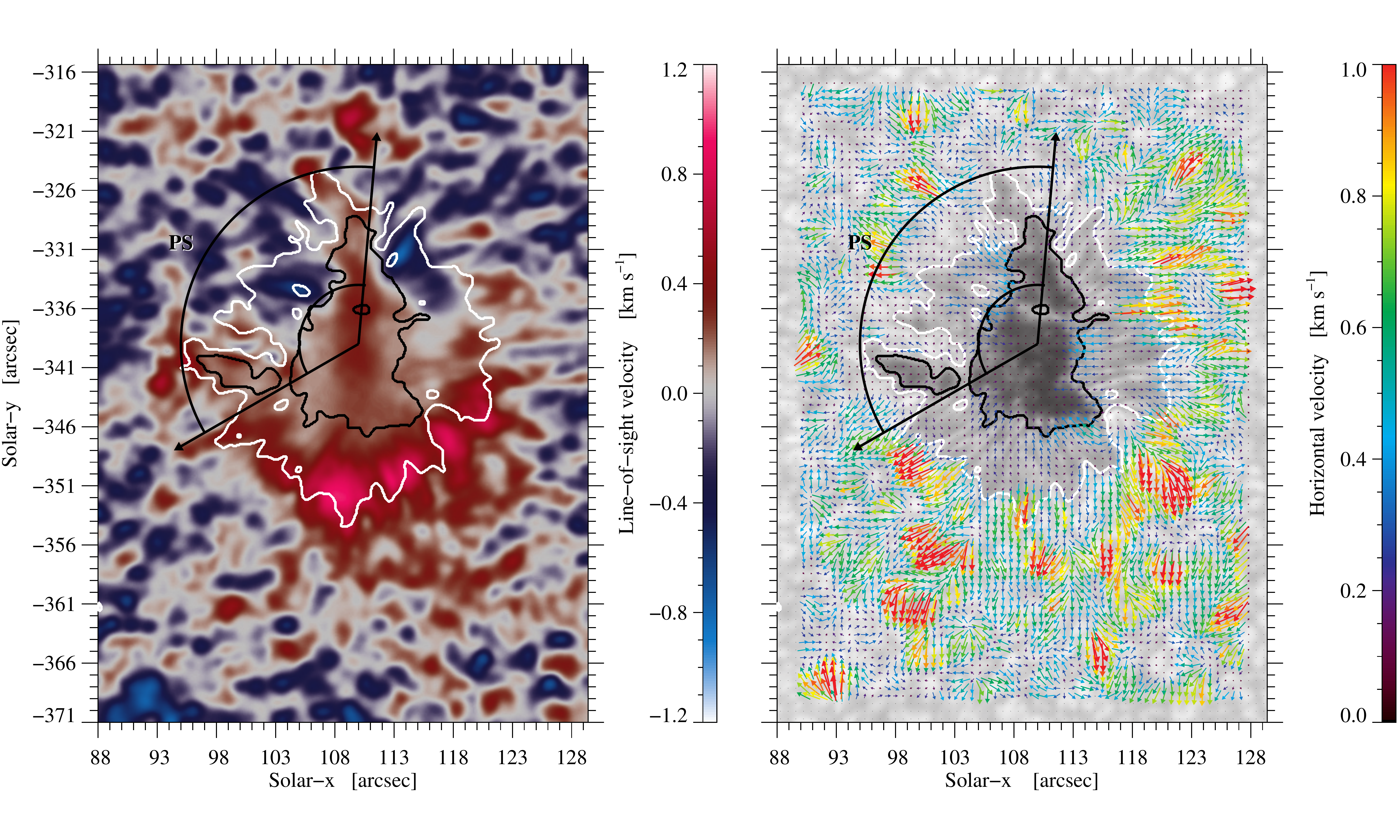}
\caption{Maps of GFPI averaged LOS velocity (\textit{left}) and horizontal proper 
    motions (\textit{right}) around the leading spot
    \textsf{P$_\mathsf{1}$}, where 
    color-coded vectors indicate magnitude and direction of the horizontal 
    flows. The black contours mark the umbra-penumbra boundary, whereas the 
    white contours denote the penumbra-granulation boundary. The contours
    shown here and in all subsequent figures are extracted from the GFPI
    broad-band image. The two black arcs mark the penumbral sector 
    \textsf{PS}.}
\label{FIG05}
\end{figure*}

The high-resolution GFPI image sequences allowed us to follow the horizontal 
proper motions around the sunspot \textsf{P$_\mathsf{1}$}. We applied LCT 
(Sect.~\ref{SEC02.2}) to the broad-band image sequences, which covered a 
40-minute time period. The average flow map is depicted as rainbow colored 
vectors in the right panel of Fig.~\ref{FIG05}. The expected global granular 
flow pattern with mesogranular cells was clearly visible within the quieter part 
of the FOV. One of the prominent flow features was a strong outward flow, i.e., 
the moat flow partly encircling \textsf{P$_\mathsf{1}$}. In addition, the 
typical divergence line \citep[e.g.,][]{Deng2007} demarcated inflows in the 
inner and outflows in the outer penumbra. However, the moat flow was absent in 
the \textsf{PS} containing the elongated umbral core. Furthermore, the overall 
proper motions were diminished in that region. Only some distance away from 
\textsf{PS}, outward motions seem to continue but with much lower flow speeds. 
The moat flow appears to terminate at the nearest supergranular boundary, which 
was already reported in earlier studies \citep[e.g.,][]{Verma2012a}.

The mean LCT velocity value over the whole FOV was $\bar{v} = 0.42 \pm 
0.26$~km~s$^{-1}$ with the maximum velocity reaching $v_\mathrm{max} = 
1.57$~km~s$^{-1}$. Furthermore, the $10^\mathrm{th}$ percentile velocity was 
$v_{10} = 0.79$~km~s$^{-1}$, and the median velocity was $v_\mathrm{med} = 
0.38$~km~s$^{-1}$. Comparing these values with previous study of  
\citet{Verma2011}, where LCT was applied to Hinode G-band images containing a 
sunspot, the current velocities are somewhat lower, which is likely caused by 
the shorter cadence $\Delta t$ in the present study. The mean velocity in moat 
was $\bar{v} = 0.49 \pm 0.27$~km~s$^{-1}$, which is in agreement with values 
presented in previous studies \citep[e.g.,][]{Sobotka2007}. 

%-------------------------------------------------------------------------------
%     Figure 5: GFPI data
%-------------------------------------------------------------------------------

%-------------------------------------------------------------------------------
%    Line-of-sight velocities
%-------------------------------------------------------------------------------

\subsection{Line-of-sight velocities}

% GFPI
Beyond the horizontal flow fields discussed in Sect.~\ref{SEC03.2}, GFPI spectra 
in the Fe\,\textsc{i} $\lambda$617.3~nm spectral line provided the photospheric 
LOS velocity in and around the leading sunspot (Sect.~\ref{SEC02.2}). The 
40-minute averaged velocity map depicted in the left panel of Fig.~\ref{FIG05} 
was scaled between $\pm 1.2$~km~s$^{-1}$. As expected the Evershed flow appears 
around the sunspot penumbra as red- and blueshifts, predominantly in the lower 
and upper parts of the spot. The Evershed flow starting in the penumbra 
continued as strong moat flow in the LCT map. The moat flow is also seen in the 
LOS velocity map, prominently at the limb-side penumbra. In that region we also 
noticed Evershed clouds \citep{CabreraSolana2006} in the time-lapse movie of LOS 
velocity maps. However, in the \textsf{PS} that faces the continuous flux 
emergence, the flows deviated from the usual penumbral flow pattern. The gap, 
which was created by the vanished penumbral filaments, displayed a flow pattern 
resembling that of granulation. The elongated umbral core in the \textsf{PS} 
possessed a velocity pattern similar to that of the umbra of sunspot 
\textsf{P$_\mathsf{1}$}. However, the tip of the elongated umbral core exhibited 
a strong redshift, while the LCT velocity remained relatively low.

\begin{figure*}
\center
\includegraphics[width=\textwidth]{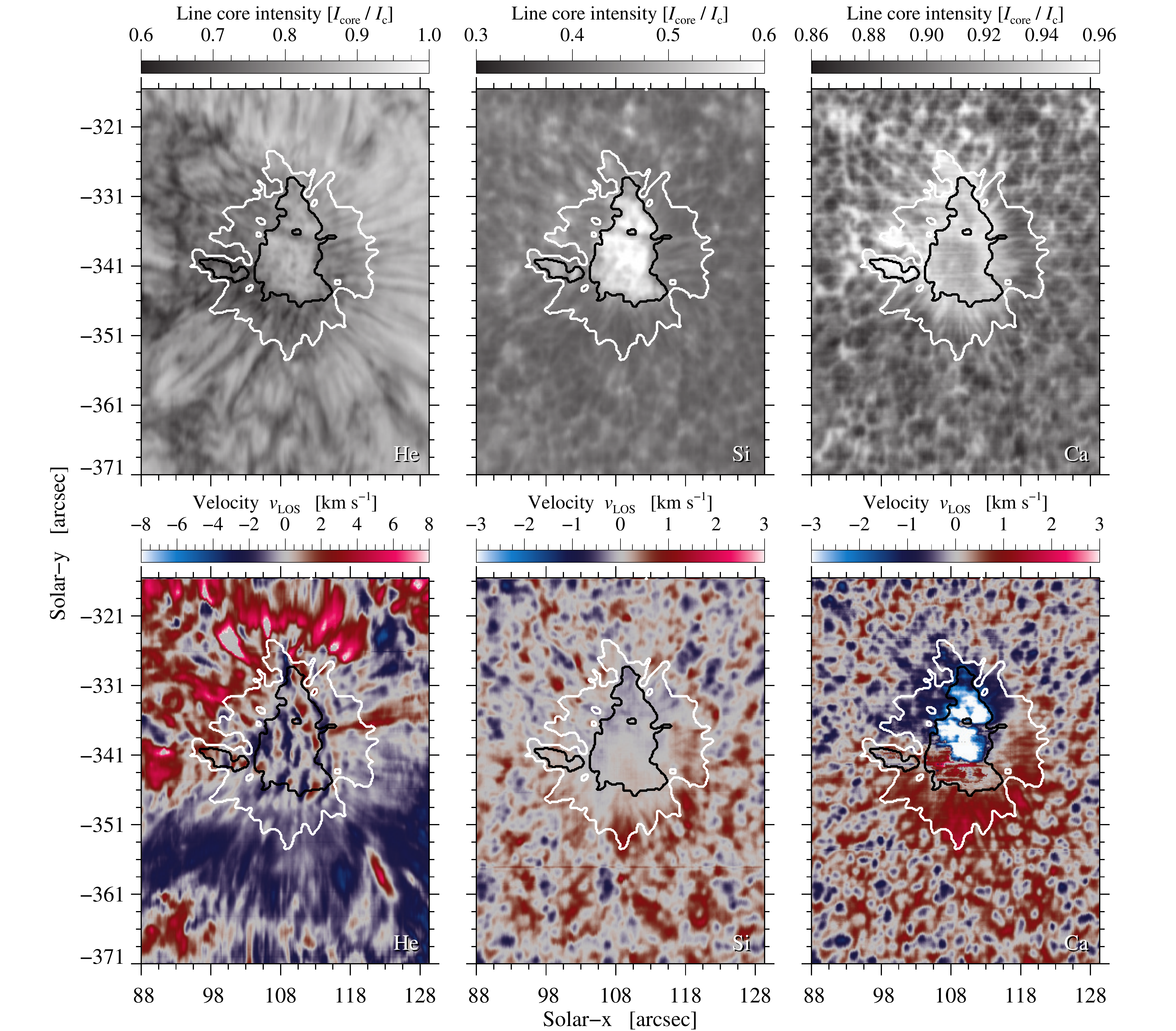}
\caption{Slit-reconstructed maps of the line-core intensity (\textit{top}) 
    and the LOS velocity (\textit{bottom}) for the spectral lines 
    He\,\textsc{i} 
    (\textit{left}), Si\,\textsc{i} (\textit{middle}), and Ca\,\textsc{i} 
    (\textit{right}) observed with GRIS at 09:02~UT on 2016 September~24.
    The contours are same as shown in Fig.~\ref{FIG05}. Simple
    line-fitting did not deliver proper fits for the Ca\,\textsc{i} line
    profiles in the umbra, which appear as  conspicuous white patches in 
    the respective velocity map.}
\label{FIG06}
\end{figure*}

% GRIS
The GRIS spectra offered a possibility to infer both photospheric and 
chromospheric LOS velocities. Figure~\ref{FIG06} depicts line-core intensity and 
LOS velocity maps for the three lines observed with GRIS. In general, these 
velocities are higher than those of the GFPI spectra. The maps for the 
Si\,\textsc{i} and Ca\,\textsc{i} lines, which represent the upper and the lower 
photosphere, respectively, show the granulation pattern as well as signatures of 
the Evershed flow. The inverse Evershed flow is visible in the He\,\textsc{i} 
map. A faint positive polarity patch, which is seen in the HMI magnetograms 
above \textsf{P$_\mathsf{1}$}, appeared as a bright spot at coordinates 
(93\arcsec, $-321$\arcsec) in the Ca\,\textsc{i} line-core images and exhibited 
strong blueshifts. Velocities for the chromospheric He\,\textsc{i} line were 
higher than photospheric velocities. As expected the chromospheric 
He\,\textsc{i} map displayed a more filamentary structure. However, the 
He\,\textsc{i} line-core intensity did not show any (super)penumbral filaments 
above \textsf{PS} showing a very perturbed penumbral structure. Bright features 
in the Ca\,\textsc{i} line-core intensity near \textsf{PS} exhibited blueshifts. 
However, the decaying sunspot \textsf{P$_\mathsf{1}$} in the Si\,\textsc{i} LOS 
velocity map had very low velocities. In addition, strong downflows are present 
in the He\,\textsc{i} map, where the \textsf{PS} faces the \textsf{EAR}, but 
there is also significant asymmetry in coverage related to the inverse Evershed 
flow. In the Ca\,\textsc{i} LOS velocity map, the Lorentzian line-core fits 
failed in some parts of the umbra of \textsf{P$_\mathsf{1}$}, because the 
corresponding spectral line profiles were very shallow and significantly 
broadened by the strong magnetic field in the umbra.

% ------------------------------------------------------------------------------
%    Figure 6: Velocity Maps GRIS
% ------------------------------------------------------------------------------

%-------------------------------------------------------------------------------
%    Line-of-sight magnetic field
%-------------------------------------------------------------------------------

\subsection{Line-of-sight magnetic field}

\begin{figure*}
\center
\includegraphics[width=\textwidth]{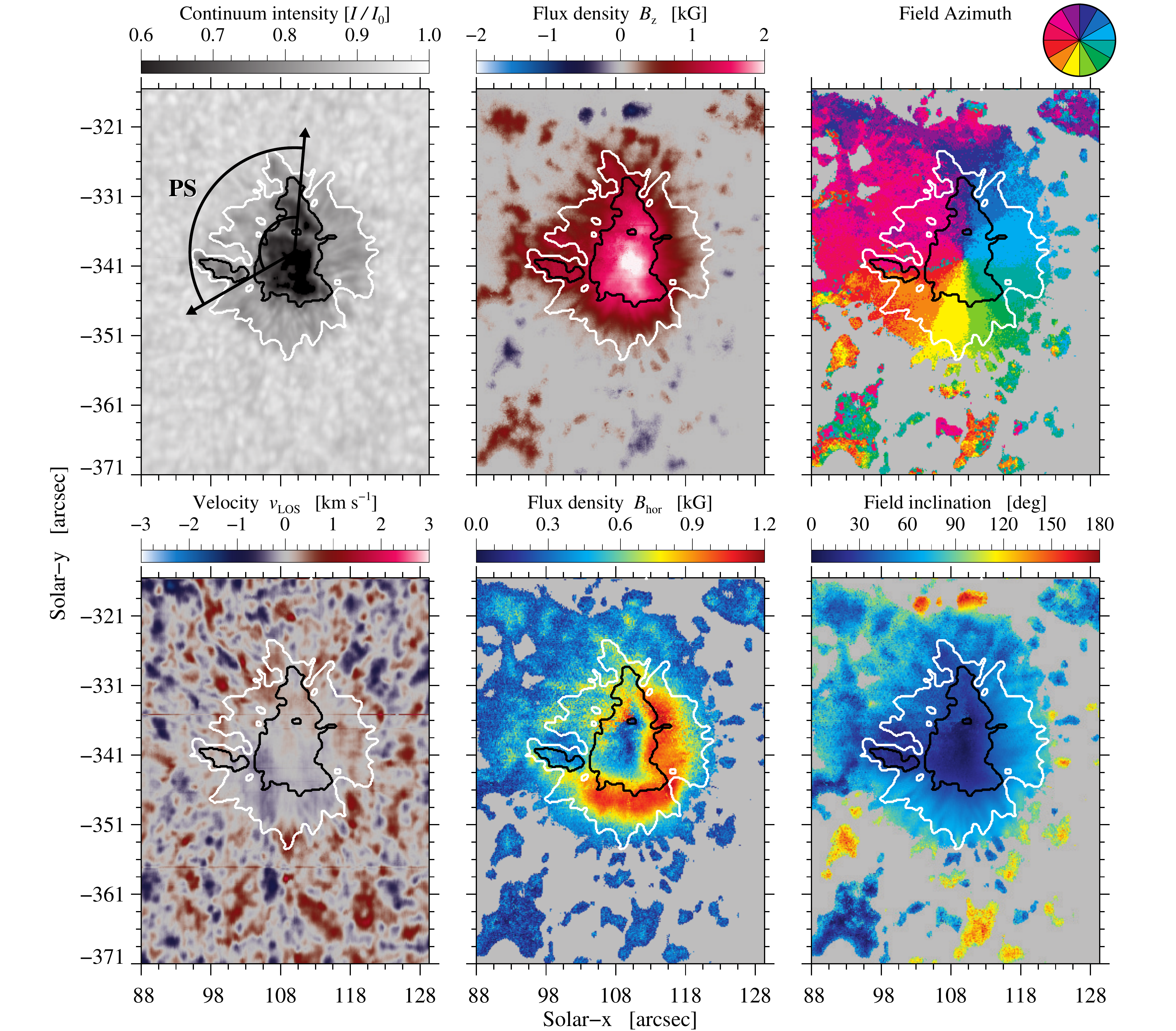}
\caption{Maps of physical parameters derived for the Si\,\textsf{i} line using 
    the SIR code for the GRIS scan starting at 09:02~UT on 2016 September~24:  
    normalized intensity $I / I_0$, vertical component of magnetic flux density
    $B_\mathrm{z}$, magnetic field azimuth $\phi$, Doppler velocity
    $v_\mathrm{LOS}$, horizontal component of magnetic flux density
    $B_\mathrm{hor}$, and magnetic field inclination $\gamma$ (\textit{top-left
    to bottom-right}). The polarization signal in the grey region is below
    the noise level. The two black arcs in the normalized intensity map 
    mark the penumbral sector \textsf{PS}. Note that $B_\mathrm{z}$, $\phi$,
    and $\gamma$ are in the local reference frame.}
\label{FIG07}
\end{figure*}

\begin{figure*}
\center
\includegraphics[width=\textwidth]{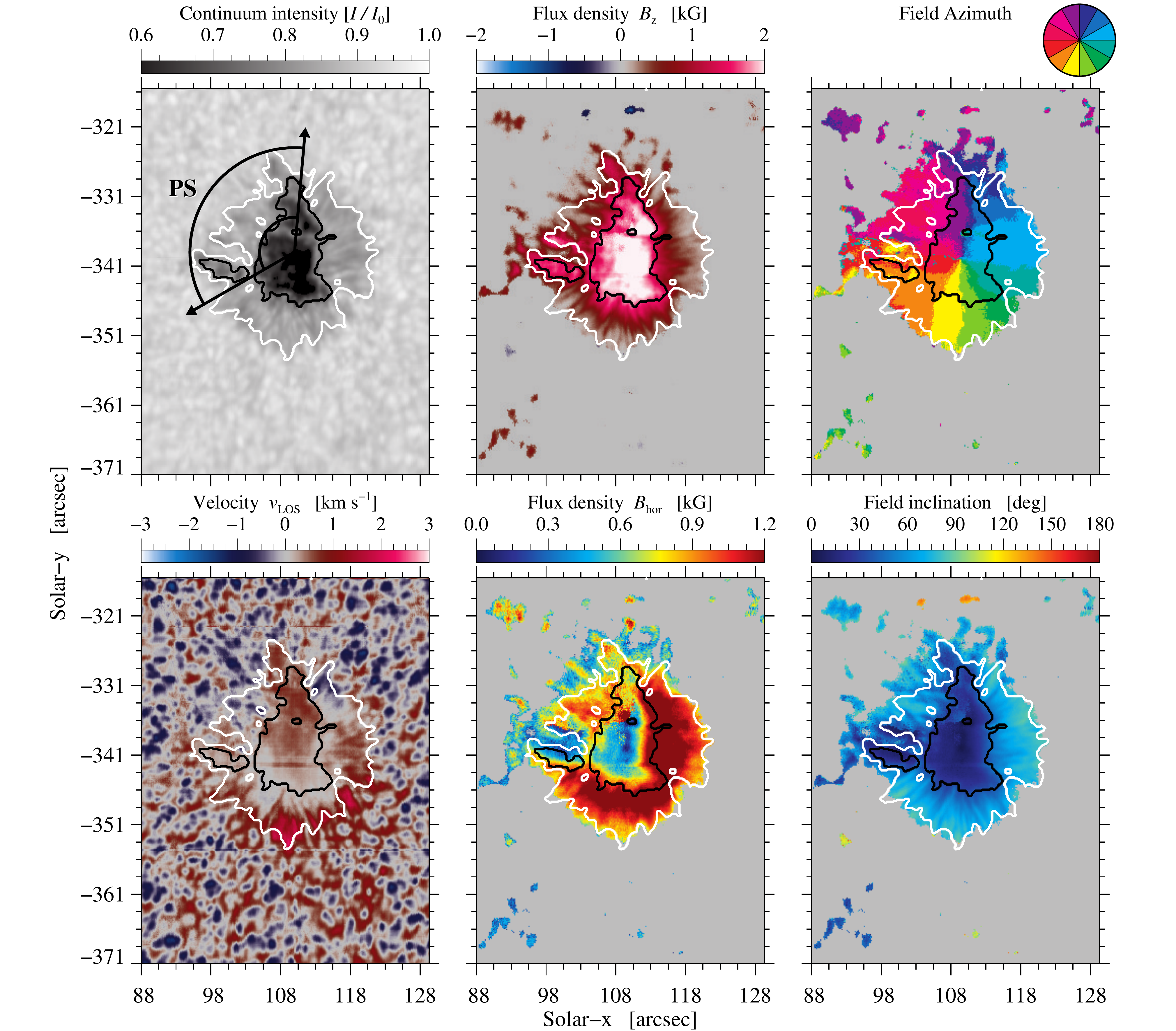}
\caption{Same as Fig.~\ref{FIG07} but parameters derived for the Ca\,\textsc{i} 
line using the SIR code for the GRIS scan starting at 09:02~UT on 2016 
September~24.}  
\label{FIG08}
\end{figure*}

% Magnetic field strength
Magnetograms obtained with HMI give an overall impression of the magnetic field 
in and around the whole active region. To scrutinize the detailed magnetic 
structure of the leading spot, we used GRIS scans. We inverted the observed 
full-Stokes spectra with the SIR code (Sect.~\ref{SEC02.2}). The inversion were 
carried out for both the photospheric Si\,\textsc{i} and Ca\,\textsc{i} lines 
and for both scans, giving us an opportunity to study the temporal evolution and 
the height dependence of the magnetic field. The physical parameters obtained 
from the Si\,\textsc{i} and Ca\,\textsc{i} inversions are displayed in 
Figs.~\ref{FIG07} and \ref{FIG08}, respectively. Since the results for both 
scans differed only in detail, only the inferred parameters from the first scan 
are shown for both lines. However, for both scans the mean values of 
$B_\mathrm{z}$, $B_\mathrm{hor}$, and inclination $\gamma$ for umbra, elongated 
umbra in \textsf{PS}, \textsf{PS} without elongated umbra, and the remainder of 
the penumbra were compiled in Table~\ref{TAB01}.

% properties in Si Line
The vertical magnetic flux density $B_\mathrm{z}$ was positive for the leading 
spot \textsf{P$_\mathsf{1}$} as seen in the Si\,\textsc{i} line 
(Fig.~\ref{FIG07}). Small negative polarity patches were found in the 
surroundings of \textsf{P$_\mathsf{1}$} in both lines. On smaller spatial 
scales, MMFs were present in the vicinity of the spot. They were mainly of 
type~II (unipolar with the same polarity as the spot) with a few type~III 
(unipolar with opposite polarity as the spot) and type~I MMFs (bipolar with the 
inner foot-point of opposite polarity as the spot) interspersed. As noted in 
Table~\ref{TAB01}, the vertical component of the magnetic field for the 
elongated umbral core in the \textsf{PS} was higher than the rest of penumbra 
and \textsf{PS}, whereas the horizontal component was lowest for the elongated 
umbral core. The values of $B_\mathrm{hor}$ for the \textsf{PS} facing the EAR 
was lower than rest of penumbra, but it contained some small-scale patches with 
high values. Even though barely noticeable in the continuum images, 
$B_\mathrm{hor}$ had a significant magnetic flux density, which extended towards 
the EAR. The field inclination for the elongated umbral core in the \textsf{PS} 
was very similar to that of the umbra, i.e., almost vertical (inclination of 
about $26^\circ$), and its structure can be easily traced in the map. For the 
rest of the penumbra, the horizontal flux density was high $B_\mathrm{hor} = 
0.75 \pm 0.22$~kG, the Evershed flow was strong, and the field inclination was 
around 50--70$^\circ$ with the typical appearance of an ``uncombed'' penumbra. 

% properties in Ca Line
The physical maps from the first scan of the Ca\,\textsc{i} line were compiled 
in Fig.~\ref{FIG08}. Since the Ca\,\textsc{i} line originates lower in the 
photosphere than the Si\,\textsc{i} line, $B_\mathrm{z}$ had a smaller areal 
extent with only few signs of MMFs. The MMFs are smaller in the $B_\mathrm{z}$ 
map. The vertical and horizontal magnetic field components had higher values in 
Ca\,\textsc{i} line than Si\,\textsc{i} line. The elongated umbral core in the 
\textsf{PS} possessed a strong $B_\mathrm{z}$ in the Ca\,\textsc{i} line as 
well. The lowest values are just above the gap intruding into the \textsf{PS}, 
which also had low $B_\mathrm{hor}$ values. However, the upper part of the 
\textsf{PS} enclosed a patch with higher $B_\mathrm{hor}$ values, but still 
lower than rest of the penumbra. The field inclination of the elongated umbral 
core in the \textsf{PS} was around $26^\circ$ in the Ca\,\textsc{i} line, with a 
distinct boundary setting it apart from the rest of the penumbra. The expansion 
of $B_\mathrm{hor}$ was smaller compared to the Si\,\textsc{i} map. Similar to 
the Si\,\textsc{i} line, the Ca\,\textsc{i} map also contained very high 
$B_\mathrm{hor}$ values for the rest of the penumbra. The field inclination for 
\textsf{P$_\mathsf{1}$} had a similar structure as in the Si\,\textsc{i} map 
with mean of around $20^{\circ}$ for the umbra.

% properties of Scan II
The inversion results and the magnetic properties of second scan (not shown) 
were virtually identical to those of the first scan for both spectral lines. 
Although, the maps for the second scan encompassed a smaller FOV, and the seeing 
was not as good. In the continuum images for both lines clockwise rotation of 
\textsf{P$_\mathsf{1}$} was apparent. The three types of MMFs were still present 
around the leading spot \textsf{P$_\mathsf{1}$}. However, they became smaller as 
compared to the previous scan. The magnetic structures in $B_\mathrm{z}$ and 
$B_\mathrm{hor}$ evolved little over the two hours. In the Si\,\textsc{i} maps, 
the elongated umbral core retained low values $B_\mathrm{hor} = 0.48 \pm 
0.14$~kG but with increased values $B_\mathrm{z} = 1.02 \pm 0.96$~kG. A similar 
change in the magnetic field properties was also observed in the Ca\,\textsc{i} 
maps for the elongated umbral core with mean values $B_\mathrm{hor} = 0.75 \pm 
0.22$~kG and $B_\mathrm{z} = 1.43 \pm 0.20$~kG.

%-------------------------------------------------------------------------------
%     Figure 07: SIR inversion results
%-------------------------------------------------------------------------------

% discussion about velocities
In addition, the LOS velocities derived with the SIR code covered the range $\pm 
3$~km~s$^{-1}$. The velocity maps of the Si\,\textsc{i} line showed some 
indications of the Evershed effect, which was clearly seen in the Ca\,\textsc{i} 
line. The structures in the LOS velocity maps were very similar to the ones 
computed by simple line-core fitting (Fig.~\ref{FIG06}). However, the LOS 
velocities derived with the SIR code for the Ca\,\textsc{i} line slightly 
differed from the ones derived with line fitting. The inversion provided 
reasonable fits for the shallow and broad line profiles of the umbra. The FOV of 
the GRIS observations contains only the leading spot \textsf{P$_\mathsf{1}$}, 
and it was possible to see the superpenumbra around it in He\,\textsc{i} line 
core. As expected the observed $Q$-, $U$-, and $V$-profiles show the extended 
magnetic field in chromosphere (not shown here), but the small FOV was 
insufficient to comment on how the overlying chromospheric canopy affected the 
decaying penumbra.

%-------------------------------------------------------------------------------
%     Figure 08: SIR inversion results
%-------------------------------------------------------------------------------

\begin{SCfigure*}
\includegraphics[width=0.65\textwidth]{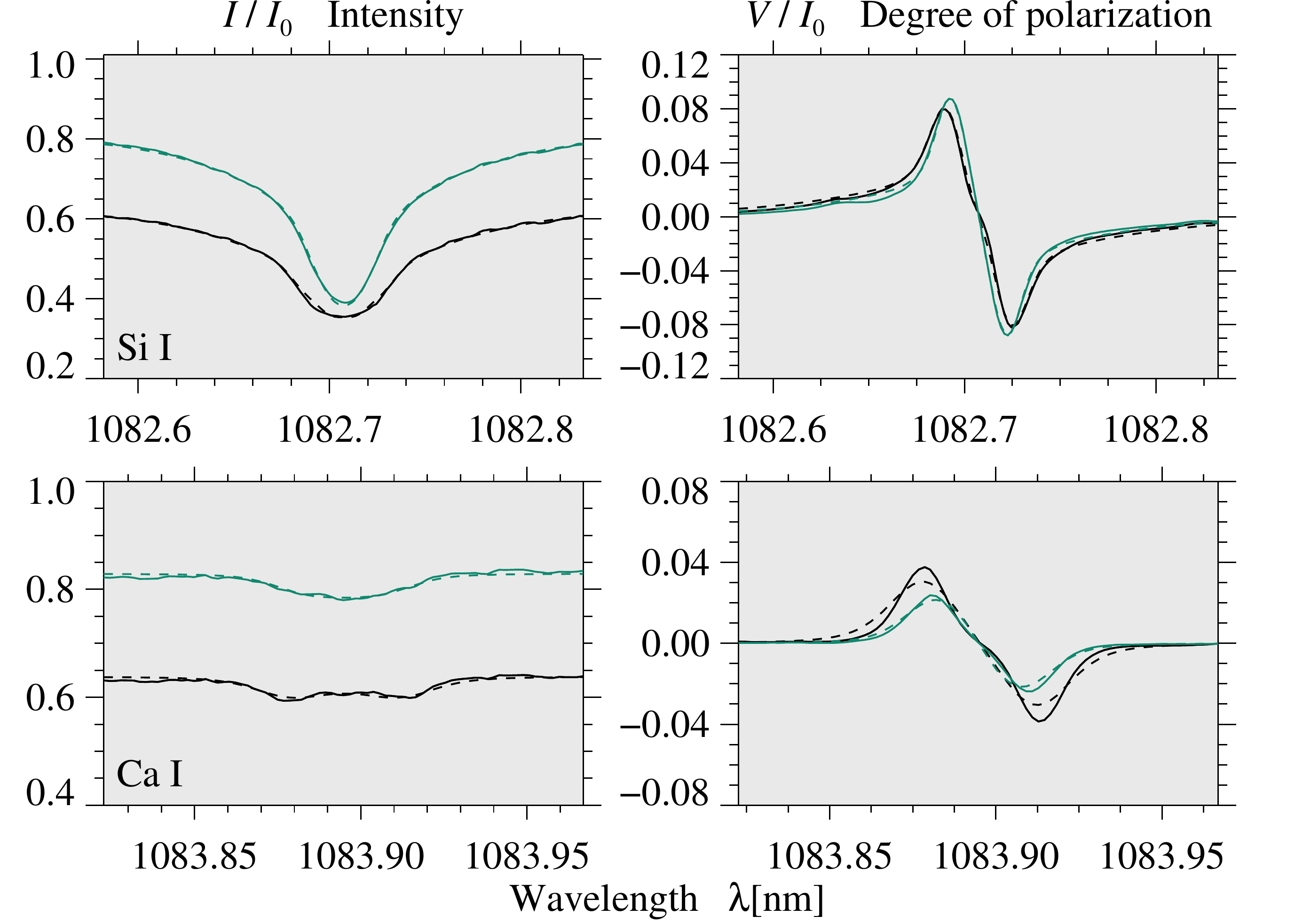}
\caption{Average Stokes-$I$ and -$V$ profiles
    (\textit{solid}) for the elongated umbral core (\textit{green}) and 
    the umbra of \textsf{P$_\mathsf{1}$} (\textit{black}) for the 
    Si\,\textsc{i} (\textit{top}) and Ca\,\textsc{i} (\textit{bottom})
    lines. The SIR results for these profiles are shown in same colors 
    but as dashed lines.}
\label{FIG09}
\end{SCfigure*}

% Comparison of values between elongated umbra and sunspot Umbra
The elongated umbral core in the \textsf{PS} has properties similar to the 
sunspot umbra. The average Stokes-$I$ and $V$-profiles in both the 
Si\,\textsc{i} and Ca\,\textsc{i} lines for the elongated umbral core and the 
\textsf{P$_\mathsf{1}$} umbra are displayed in Fig.~\ref{FIG09}. The intensity 
of the elongated umbral core exceeded that of the umbra. However, Stokes-$V$ 
exhibited a similar degree of polarization in both lines. The mean B$_z$ in 
Si\,\textsc{i} line for the elongated umbral core was $0.99 \pm 0.12$~kG and was 
lower in comparison to $1.62 \pm 0.23$~kG the for umbra. These values were 
higher in the Ca\,\textsc{i} line for both features, with $1.30 \pm 0.27~$kG for 
the elongated umbral core and $1.94 \pm 0.22$~kG for the umbra. The average 
inclination for the elongated umbral core was $26.2 \pm 8.8^\circ$ and $27.8 \pm 
13.6^\circ$ in the Si\,\textsc{i} and Ca\,\textsc{i} lines, respectively. For 
the umbra the mean inclination were $21.2 \pm 8.3^\circ$ and $20.2 \pm 
7.9^\circ$ in the Si\,\textsc{i} and Ca\,\textsc{i} lines, respectively. 
The values for B$_z$ in the elongated umbral core 
increased from the first to the second scan for both spectral lines. However, 
as mentioned, the seeing was not good during the second scan, which could lead 
to a higher contribution of stray-light. Additonally, the 
used inversion scheme does not account for the changing seeing conditions. 
Hence, the direct comparison of absolute values of the magnetic field are 
diffcult. However, in comparison, the elongated umbral core has properties 
resembling those of the umbra in both scans, but they are not entirely 
identical. The values of B$_z$ in Ca\,\textsc{i} for the elongated umbral core 
do reach the values as seen by \citet{Jurcak2015} during the penumbra formation. 
However, we do not have simultaneous Hinode data to confirm whether it fulfills 
the criteria proposed by \citet{Jurcak2011} to establish a stable 
umbra-penumbra boundary.

%-------------------------------------------------------------------------------
%     Table I
%-------------------------------------------------------------------------------

%-------------------------------------------------------------------------------
%    Table 1 (Magnetic field properties for various features )
%-------------------------------------------------------------------------------
\begin{table}[t]
\footnotesize
\caption{Average magnetic field properties derived from SIR for umbra, 
    elongated umbral core, penumbral sector \textsf{PS}, and the remainder 
    of the penumbra for both the Si\,\textsc{i} and the Ca\,\textsc{i} 
    spectral lines and for both GRIS scans.}
\phn
\begin{tabular}{lccc}
\hline\hline
Features                                   & $B_\mathrm{z} \pm \sigma_{B_\mathrm{z}}$
                                           & $B_\mathrm{hor} \pm \sigma_{B_\mathrm{hor}}$
                                           & $\gamma \pm \sigma_{\gamma}$ \rule[-1.5mm]{0mm}{5mm}\\
                                           & [kG] & [kG] & [deg] \rule[-1.5mm]{0mm}{5mm}\\ 
\hline
                                           &                    &     Si\,\textsc{i} -- Scan I       &    
                                            \rule[0mm]{0mm}{3.5mm}\\
                                            
                                           \cmidrule[0.4pt]{2-4}
Umbra                                      &  1.62 $\pm$ 0.23   &  0.61 $\pm$ 0.22   & 21.25 $\pm$  \phn8.38 \\
Elongated Umbra                            &  0.99 $\pm$ 0.12   &  0.48 $\pm$ 0.15   & 26.15 $\pm$  \phn8.85 \\
Penumbral Sector                           &  0.83 $\pm$ 0.24   &  0.57 $\pm$ 0.15   & 35.31 $\pm$  10.30 \\
Penumbra                                   &  0.59 $\pm$ 0.42   &  0.75 $\pm$ 0.22   & 56.80 $\pm$ 14.53 \\
                                           \cmidrule[0.4pt]{2-4}
                                           &                    &     Si\,\textsc{i} -- Scan II       &    
                                           \rule[0mm]{0mm}{3.5mm}\\ 
                                           \cmidrule[0.4pt]{2-4}
Umbra                                      & 1.53 $\pm$ 0.20    & 0.65 $\pm$ 0.22    & 23.29 $\pm$ \phn8.60  \\
Elongated Umbra                            & 1.02 $\pm$ 0.10    & 0.48 $\pm$ 0.13    & 24.99 $\pm$ \phn6.62  \\
Penumbral Sector                           & 0.76 $\pm$ 0.24    & 0.55 $\pm$ 0.16    & 37.04 $\pm$ 11.50 \\
Penumbra                                   & 0.61 $\pm$ 0.41    & 0.70 $\pm$ 0.21    & 54.50 $\pm$ 14.96 \\                                                                                                   \cmidrule[0.4pt]{2-4}
                                           &                    &     Ca\,\textsc{i} -- Scan I        &     
                                           \rule[0mm]{0mm}{3.5mm}\\ 
                                           \cmidrule[0.4pt]{2-4}
Umbra                                      & 1.94 $\pm$ 0.22    & 0.71 $\pm$ 0.27    & 20.20 $\pm$ \phn7.86  \\
Elongated Umbra                            & 1.30 $\pm$ 0.27    & 0.69 $\pm$ 0.27    & 27.80 $\pm$ 13.65  \\
Penumbral Sector                           & 0.94 $\pm$ 0.43    & 0.74 $\pm$ 0.22    & 40.81 $\pm$ 17.62  \\
Penumbra                                   & 0.61 $\pm$ 0.50    & 1.05 $\pm$ 0.24    & 64.57 $\pm$ 16.38  \\                                                                                                  \cmidrule[0.4pt]{2-4}                                           
                                           &                    &     Ca\,\textsc{i} -- Scan II        &    
                                           \rule[0mm]{0mm}{3.5mm}\\ 
                                           \cmidrule[0.4pt]{2-4} 
Umbra                                      & 1.87 $\pm$ 0.22    & 0.78  $\pm$ 0.29   & 22.77 $\pm$ \phn9.17  \\
Elongated Umbra                            & 1.43 $\pm$ 0.20    & 0.75  $\pm$ 0.22   & 27.68 $\pm$ \phn9.89  \\
Penumbral Sector                           & 0.80 $\pm$ 0.40    & 0.83  $\pm$ 0.22   & 48.62 $\pm$ 17.41 \\
Penumbra                                   & 0.63 $\pm$ 0.51    & 1.02  $\pm$ 0.24   & 63.16 $\pm$ 18.31  \\ 
\hline                                            
\end{tabular}
\label{TAB01}
\end{table}

%-------------------------------------------------------------------------------
%     Figure 9: Averaged I and V profiles
%-------------------------------------------------------------------------------

\begin{SCfigure*}
\includegraphics[width=0.65\textwidth]{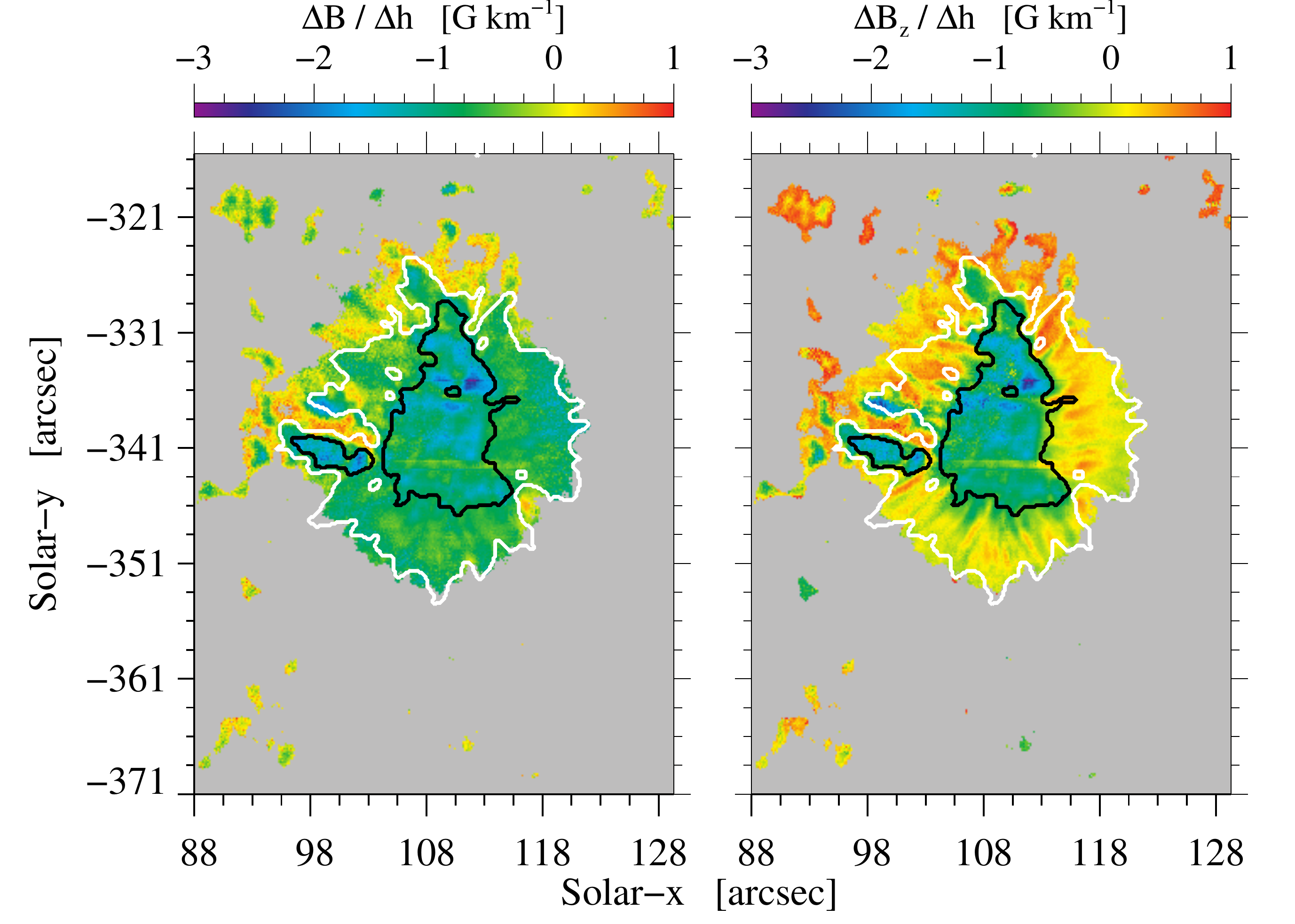}
\caption{Difference of the total magnetic field 
    strength (\textit{left}) and vertical component of the magnetic field 
    (\textit{right}) from the spectral lines Si\,\textsc{i} and Ca\,\textsc{i},
    respectively, divided by the height difference.}
\label{FIG010}
\end{SCfigure*}

% Discussion about height dependence
To study the height dependence of the magnetic field, we divided the difference 
of total $B$ and vertical component $B_\mathrm{z}$ of the magnetic flux density 
for the Si\,\textsc{i} and Ca\,\textsc{i} lines by the height difference of both 
lines for the first scan. The height for both lines was computed according to 
the method described in \citet{Balthasar2008} using the temperature map for both 
lines at $\tau = 1$ obtained via SIR and contribution functions from two model 
atmospheres, i.e., an umbral model \citep[M4,][]{Kollatschny1980} and a 
quiet-Sun model \citep[t93--27,][]{Schleicher1976}. The Si\,\textsc{i} line 
originates roughly about 350~km higher in the atmosphere than the Ca\,\textsc{i} 
line. The computed difference maps are compiled in Fig.~\ref{FIG010}.

The mean value of the $B$ gradients in the \textsf{P$_\mathsf{1}$} umbra of 
$-1.06 \pm 0.35$~G~km$^{-1}$ and in the elongated umbral core of $-1.31 \pm 
0.41$G~km$^{-1}$ was similar. The negative values indicate that the field 
strength decreases with height. The mean value for the penumbra of $-0.69 \pm 
0.33$~G~km$^{-1}$ was lower compared to both umbra and elongated umbral core. At 
the penumbra-granulation border, the field strength increased with height. In 
the gradient map for $B_\mathrm{z}$, the trend for umbra and elongated umbral 
core was similar to the map for $B$. The mean value for the $B_\mathrm{z}$ 
gradient was negative in both the umbra of $-0.96 \pm 0.42$~G~km$^{-1}$ and the 
elongated umbral core $-1.12 \pm 0.50$~G~km$^{-1}$, which even slightly exceeded 
the umbral gradient. However, the gradient the for penumbra had small negative 
values with a mean of $-0.09 \pm 0.42$~G~km$^{-1}$. The penumbral sector 
\textsf{PS} had a few patches with negative gradients. However, it also 
contained a positive gradient patch at the penumbra-granulation boundary. The 
magnetic field is less inclined higher in the atmosphere. Hence, the vertical 
component increases with the height leading to positive gradient values. In both 
gradient, maps the umbra and elongated umbral core exhibited a similar trend.

%===============================================================================
%    Discussion
%===============================================================================

\section{Discussion and conclusions}

How and when does a sunspot decay? Does the penumbra simply disappear and 
do changes occur in magnetic field properties spontaneously? When the GREGOR
observations started, the leading spot began to decay, which became evident,
when two light-bridges appeared in the umbra. Light-bridges are an indication 
of impending sunspot fragmentation \citep{Vazquez1973}. One of the 
light-bridges disappeared and an elongated, dark umbral core formed at its 
edge within the \textsf{PS}. The penumbral filaments in this sector did 
not simply vanish but formed dark features with umbra-like velocity and 
magnetic field properties. Light-bridges are suggested as a trigger for 
penumbra formation by \citet{Rezaei2012}. However, in our case the 
light-bridge initiated the transformation of penumbral filaments into an
elongated, dark umbral core. This conversion ultimately led to the 
disappearance of the penumbral filaments at this location.

% Relation between horizontal and LOS velocities.
With the GREGOR telescope and its instruments, we have access to the
three-dimensional flow field in and around the leading sunspot. The usual 
outward directed moat flow was present around the spot in horizontal flow
maps. However, it was absent within the \textsf{PS} with the elongated, 
dark umbral core. Here, the LCT velocities were low indicating constrained
horizontal plasma motions. The photospheric GFPI and GRIS LOS velocity maps
reveal the Evershed flow around the sunspot. However, this distinct flow 
pattern was absent in the above mentioned \textsf{PS}. Signatures of 
convective motions were noticeable in the LOS velocity maps in the region 
between the elongated, dark umbral core and the neighboring penumbral 
filaments. The umbral core displayed low LOS velocities with an average 
velocity near zero. This indicates that the elongated, dark umbral core
appearing in place of penumbral filaments had flow properties similar 
to an umbra in both horizontal and LOS velocities maps.

%-------------------------------------------------------------------------------
%     Figure 10: Difference maps
%-------------------------------------------------------------------------------

% Magnetic field properties
The magnetic field properties of the decaying \textsf{PS} differed from the 
usual penumbra. The elongated, dark umbral core possessed a low horizontal 
field component, but a strong vertical magnetic field component. Furthermore, 
the magnetic field inclination was also close to zero. This indicated that 
the field lines in this location of penumbra were nearly vertical and that 
the extruding umbral core resembled a typical sunspot umbra. The magnetic 
field properties extracted from the Si\,\textsc{i} and Ca\,\textsc{i} lines
were slightly different but followed a similar trend. The vertical and 
horizontal magnetic field components in the lower photosphere (Ca\,\textsc{i}
line) were more confined and stronger. In the spectropolarimetic observations 
of a decaying sunspot \citet{BellotRubio2008} found small-scale 
inhomogeneities in the surrounding of a naked umbra. These finger-like 
features had weak and nearly horizontal magnetic field. They suggested that
the penumbra at photospheric heights will disappear, when the penumbral 
field lines, which no longer carry strong Evershed flows, rise to the
chromosphere. In our data, even-though not all penumbral filaments vanished, 
we noticed that the \textsf{PS} facing the flux emerging site did not display 
the usual Evershed and moat flow. In addition, the horizontal magnetic field 
was weaker for the entire \textsf{PS}. Although, the rest of the 
sunspot penumbra showed the typical flow and magnetic field properties, 
including the presence of small-scale moving magnetic features.

% rapid penumbra decay
Rapid penumbral decay is closely related to flares \citep{Wang2004, Deng2005}
and much faster (a few minutes) than the usual decay process.
\citet{Beauregard2012} observed penumbral decay and umbral strengthening after 
an X-class solar flare. For the same solar flare, \citet{WangS2012} 
investigated the response of the photospheric magnetic field. They found
evidence of a rapid and irreversible enhancement of the horizontal magnetic 
field at the flaring magnetic polarity inversion line. They related the
strengthening  of horizontal magnetic field to newly formed low-lying fields
resulting from tether-cutting reconnection \citep{Moore2001}. Rapid
rearrangement of penumbral magnetic field lines in flaring $\delta$-sunspots 
was the focus of the work by \citet{Wang2012}. They suggested that the magnetic 
field became more vertical due to restructuring associated with flares and
concluded that this led to a transformation of dark penumbral filaments to
faculae. In the present study, we found a similar restructuring of the 
horizontal component of the magnetic field and a strengthening of the 
vertical magnetic field component as well as the appearance of an elongated, 
dark umbral core in \textsf{PS}. However, the duration of the transformation 
took place on much longer time-scales than those inferred by 
\citet{Wang2012}. Thus, also slow changes of the magnetic field topology 
within an active region, e.g., introduced by flux emergence, can induce the
transformation of penumbral to umbral magnetic fields.

% Role played by neighboring flux emergence
How does the surrounding magnetic field affect sunspot structure and evolution? 
Looking just beyond the FOV covered by GREGOR observations, we noticed that the 
decaying penumbral sector faced a region of continuous flux emergence (EAR). 
Recently, \citet{Murabito2017} observed the formation of stable penumbra at the 
site facing the flux emergence. Their observations agree with the scenario 
proposed by \citet{Lim2013}, which suggests the pre-existing chromospheric 
canopy field assists in the formation of the penumbral filaments in the flux 
emergence region. However, in our case the penumbral sector facing flux 
emergence site decayed. The side of the sunspot facing the flux emergence site 
was stretched out towards the EAR.  In addition, the 
approaching polarity of the EAR was the same as of the decaying sunspot. The 
same polarity flux might have influenced the magnetic cannopy of the decaying 
sunspot disrupting the stable horizontal penumbral field \citep{Kuenzel1969}. 

In time-lapse movies of HMI continuum images, HMI magnetograms, and AIA 
$\lambda$171~nm images, we observed that the leading polarity of the EAR 
continuously pushed the elongated, dark umbral core and the whole \textsf{PS} of 
the leading spot \textsf{P}$_\mathsf{1}$, resulting in rotation of the entire 
sunspot. In the sketch represented inFig.~\ref{FIG04}, we summarized these 
dynamical properties seen in the region. \citet{Botha2011} studied the decay 
process of large magnetic flux tubes, such as sunspots, on a supergranular 
scale. Their nonlinear cylindrical simulation consisted of a well-defined 
central flux tube and an annular convection cell surrounding it. They found that 
the evolving azimuthal nonlinear convection in the annular cells breaks it, and 
the magnetic flux from the central flux tube then slips between these fragmented 
cells. This leads to a reduction of magnetic pressure and an increase of 
convection in the central flux tube. The convection inside the tube grows until 
the spot starts to  disintegrate. Hence, they concluded that the decay of 
central flux tube depends on the convection in the surroundings, which can add 
or remove flux from the central tube. In addition, convection in the vicinity of 
the spot can change the shape of the central flux tube. Our decaying sunspot was 
in a vicinity of a newly emerging flux system, where the separating polarities 
of the EAR pushed and rotated the DAR. Our observations imply that the decay of 
penumbra involved interaction of two flux systems, which changed the surrounding 
motions. We conclude that the continuous flux emergence in the region likely 
altered the convection pattern in the surrounding, resulting in an accelerated 
decay of penumbral filaments and led to significant rotation of the leading 
sunspot in the DAR.

% Penumbra formation around flux emergence region
The relation between penumbra formation and the surrounding flux emergence 
remains contradictory. \citet{Schlichenmaier2010b, Schlichenmaier2012} and 
\citet{Rezaei2012} found that a sunspot penumbra only develops in a region with 
stable conditions. In their studies, the side of the sunspot facing the flux 
emerging site did not develop a penumbra. However, \citet{Lim2013} and 
\citet{Murabito2017} found penumbra formation on the side of the sunspot facing 
the flux emerging region. Both agreed that the overlying chromospheric canopy 
plays a role in the formation of a penumbra. In our sunspot, the penumbral 
sector facing the flux emergence site was already in the decay phase at the 
start of the GREGOR observations. In addition, we did 
not perform a detailed analysis of the chromospheric magnetic field because the 
FOV was not sufficiently large. Hence, it is difficult to pin-point if it 
initiated and/or assisted the decay process. However, we can state that the 
magnetic and flow changes caused by the ongoing flux emergence did affect the 
penumbral decay process around the sunspot and likely changed the stable or 
``quiet'' conditions that are required to sustain a stable penumbra.

In the current work, we presented high-resolution spectroscopic and 
polarimetric observations of a decaying sunspot. At the time of the GREGOR
observations, the sunspot contained light-bridges indicating the start of 
its fragmentation. After the decay of one of the light-bridges, an elongated,
dark umbral core appeared at its edge within the decaying penumbral sector. 
The flow and magnetic field properties of this penumbral sector exhibited
weak Evershed flow, moat flow, and horizontal magnetic fields. Concurrently, 
a new flux system emerged in the already established flux system. The 
interaction between emerging and already established flux system played an
important role in penumbral decay and additionally resulted in rotation of the
leading sunspot. This hints at a subphotospheric interaction of both flux 
systems releasing twist and writhe in the DAR. In future work, we will analyze 
and follow the evolution of the active region based on the full set of
high-resolution data acquired with GREGOR, VTT, NST, and DST, which was
obtained on several days during the coordinated campaign. As of yet, we 
only discussed the photospheric magnetic field properties of active region 
NOAA~12597. In next step, we will include the inversion results of 
the chromospheric He\,\textsc{i} triplet. These high-resolution observations
are needed to provide a complete and comprehensive picture of the sunspot 
decay process.

%===============================================================================
%    Acknowledgements
%===============================================================================

\begin{acknowledgements}
The 1.5-meter GREGOR solar telescope was build by a German consortium under the 
leadership of the Kiepenheuer-Institut f\"ur Sonnenphysik in Freiburg with the 
Leibniz-Institut f\"ur Astrophysik Potsdam, the Institut f\"ur Astrophysik 
G\"ottingen, and the Max-Planck-Institut f\"ur Sonnensystemforschung in 
G\"ottingen as partners, and with contributions by the Instituto de 
Astrof\'{\i}sica de Canarias and the Astronomical Institute of the Academy of 
Sciences of the Czech Republic. The National Solar Observatory is operated by 
the Association of Universities for Research in Astronomy under a cooperative 
agreement with the National Science Foundation. SDO HMI and AIA data are 
provided by the Joint Science Operations Center -- Science Data Processing. MS 
is supported by the Czech Science Foundation under the grant 14-04338S. This 
study is supported by the European Commission's FP7 Capacities Programme under 
the Grant Agreement number 312495. R.R.\ acknowledges financial support by the 
Spanish Ministry of Economy and Competitiveness through project AYA2014-60476-P.
SJGM is grateful for financial support from the Leibniz Graduate School 
for Quantitative Spectroscopy in Astrophysics, a joint project of AIP and the 
Institute of Physics and Astronomy of the University of Potsdam and he 
acknowledges support of the project VEGA 2/0004/16.
\end{acknowledgements}

%===============================================================================
%    Bibliography
%===============================================================================

% 
% \bibliographystyle{aa}
% \bibliography{aa-jour,meetu}

\end{document}